\documentclass[aps,prd,twocolumn,superscriptaddress]{revtex4}

\usepackage{graphicx}         
\usepackage{amssymb}
\usepackage{amsmath}
\usepackage{amsthm}
\usepackage[colorlinks]{hyperref}
\usepackage{color,rotating}
\usepackage{bbm}
\usepackage{array}
\usepackage{pstricks}
\usepackage{pstricks-add}
\usepackage{colonequals}

\definecolor{darkgreen}{rgb}{0,0.6,0}
\definecolor{gray}{rgb}{.7,.7,.7}

\DeclareMathAlphabet{\EuRoman}{U}{eur}{m}{n}
\SetMathAlphabet{\EuRoman}{bold}{U}{eur}{b}{n}

 \graphicspath{{./figures/}}

\def\di{\displaystyle}

\def\bg{\begin{eqnarray}\begin{array}{rcl}\displaystyle}
\def\eg{\end{array} &\di    &\di   \end{eqnarray}}
\def\bm#1{\begin{eqnarray}\begin{array}{#1}\di}
\def\bmo#1{\begin{eqnarray*}\begin{array}{#1}\di}
\def\bml#1#2{\begin{eqnarray}\begin{array}{#1}\label{#2}\di}
\def\bgo{\begin{eqnarray*}\begin{array}{rcl}\displaystyle}
\def\ego{\end{array} &\di    &\di \nonumber  \end{eqnarray*}}

\def\btensor#1#2{\renew\left#1\begin{array}{#2}\di}
\def\brtensor#1#2#3{\ren#3\left#1\begin{array}{#2}}
\def\botensor#1#2{\renew\left#1\begin{array}{#2}}
\def\etensor#1{\end{array}\right#1}

\def\eq#1{(\ref{#1})}

\def\s0#1#2{\mbox{\small{$ \frac{#1}{#2} $}}}
\def\0#1#2{\frac{#1}{#2}}

\def\s{\sigma}

\def\ren#1{\renewcommand{\arraystretch}{#1}}

\def\renew{\renewcommand{\arraystretch}{1}}

\newcommand{\UV}{{\small UV}}
\newcommand{\IR}{{\small IR}}
\newcommand{\FRG}{{\small FRG}}
\newcommand{\RG}{{\small RG}}

\begin{document}

\title{Global Flows in Quantum Gravity}

%
\author{N. Christiansen}
\affiliation{Institut f\"ur Theoretische Physik, Universit\"at Heidelberg,
Philosophenweg 16, 69120 Heidelberg, Germany}
\author{B. Knorr}
\affiliation{Institut f\"ur Theoretische Physik, Universit\"at Heidelberg,
Philosophenweg 16, 69120 Heidelberg, Germany}
\author{J. M. Pawlowski}
\affiliation{Institut f\"ur Theoretische Physik, Universit\"at Heidelberg,
Philosophenweg 16, 69120 Heidelberg, Germany}
\affiliation{ExtreMe Matter Institute EMMI, GSI Helmholtzzentrum f\"ur
Schwerionenforschung mbH, Planckstr.\ 1, 64291 Darmstadt, Germany}
\author{A. Rodigast}
\affiliation{Institut f\"ur Theoretische Physik, Universit\"at Heidelberg,
Philosophenweg 16, 69120 Heidelberg, Germany}


\begin{abstract}
We study four-dimensional quantum gravity using non-perturbative
renormalization group methods. We solve the corresponding equations
for the fully momentum-dependent propagator, Newton’s coupling and the
cosmological constant. For the first time, we obtain a global phase
diagram where the non-Gaussian ultraviolet fixed point of asymptotic
safety is connected via smooth trajectories to a classical infrared fixed
point. The theory is therefore ultraviolet complete and
deforms smoothly into classical gravity as the infrared limit is approached. 
\end{abstract}

\maketitle

\section{Introduction}

Understanding the quantization of the gravitational force is an
outstanding problem in theoretical physics. Any viable theory of
quantum gravity must connect stable infrared (\IR) physics with a
well-behaved ultraviolet (\UV) limit. The asymptotic safety scenario
provides a \UV-completion in a natural way. It is based on a
non-Gaussian \UV{} fixed point, which leads to vanishing
$\beta$-functions in the limit of arbitrarily high energy scales and
renders the couplings finite even beyond the Planck scale
\cite{Weinberg:1980gg}. 

The asymptotic safety scenario received growing attention during the
past decades and has been studied with different methods.  The
underlying fixed point structure was found in the non-perturbative
continuum approach
\cite{Donkin:2012ud,Christiansen:2012rx,%
Reuter:1996cp,Litim:2003vp,Niedermaier:2006wt,Percacci:2007sz,Codello:2008vh,
Litim:2011cp, Reuter:2012id,
  Nagy:2012ef}, as
well as in lattice simulations
\cite{Hamber:2009mt,Ambjorn:2012jv,Ambjorn:2013tki}.  The former is
based on the functional renormalization group, in particular on its
formulation for the effective action \cite{Wetterich:1992yh}. The
crucial \UV{} fixed point is confirmed in various
approximations, including the coupling to gauge and matter fields
\cite{Percacci:2002ie,Daum:2010bc,Eichhorn:2011pc,Folkerts:2011jz,Harst:2011zx,
  Dona:2013qba}, dilaton gravity \cite{Henz:2013oxa} and higher
derivative calculations
\cite{Benedetti:2009rx,Falls:2013bv,Ohta:2013uca,Benedetti:2013jk,Demmel:2014}.
There is also a rich field of phenomenological applications based on
asymptotically safe quantum gravity. This includes e.g.\ implications
for the standard model and its extensions
\cite{Shaposhnikov:2009pv,Antipin:2013bya}, black hole physics
\cite{Falls:2010he,Falls:2012nd,Koch:2013owa,Koch:2014cqa}, collider
experiments \cite{Litim:2007iu,Gerwick:2011jw} and cosmology
\cite{Weinberg:2009wa,Copeland:2013vva}.  However, the standard
calculations lead to an ill-defined \IR{} limit since the
trajectories exhibit a singular behaviour on large length scales.

In \cite{Donkin:2012ud} an \IR{} fixed point has been found
as the endpoint of a singular line. The existence of this fixed point
was only seen within a proper distinction of background and dynamical
couplings. The singular behaviour in the vicinity of the \IR{} fixed
point was attributed to the approximation. The first smooth \IR{}
fixed point in asymptotically safe gravity, allowing for a theory that
is well defined on all energy scales, was found in
\cite{Christiansen:2012rx}. There, however, a non-classical behaviour
in the vicinity of the \IR{} fixed point has been computed,
therefore leading to modified gravity on very large length
scales. Again, the non-classical behaviour may be attributed to the
approximation. Further discussions on this issue from different
perspectives can be found in
\cite{Litim:2012vz,Nagy:2012rn,Rechenberger:2012pm, ContrerasLitim}.

In the present work, we construct a qualitatively enhanced
approximation within the systematic vertex expansion scheme introduced
in \cite{Christiansen:2012rx}. With this enhanced approximation the
theory is asymptotically safe in the \UV{}, and exhibits an \IR{}
fixed point which describes classical gravity. The corresponding
renormalization group (\RG) trajectories are globally smooth. They
connect the known \IR{} physics with classical gravity on large
length scales with a viable theory of Planckian and trans-Planckian
gravity. The present scenario encodes that quantum gravity effects
set in at about the Planck-scale, and are absent for energy-scales $E
\ll M_{\mathrm{Pl}}$.

This work improves the vertex expansion scheme set-up in
\cite{Christiansen:2012rx} in several aspects. We compute, for the
first time, fully momentum-dependent wave-function renormalizations
for the graviton and the ghost.  Note that the wave function
renormalizations are functions of the covariant Laplacian, and hence
this takes into account infinitely many terms in an expansion of the
effective action in powers of the covariant derivative. Additionally,
multi-graviton interactions are constructed from their scaling
behaviour, see \cite{Fischer:2009tn}: the dependence on the running
\RG{}-scale is deduced from consistent \RG{}-scaling of the vertex. This
novel self-consistent vertex construction is of major importance for
the transition from \UV{} to \IR{} scaling, and the stability of the
\IR{} regime.

The present work is organized as follows: We introduce our
approximation scheme in \autoref{sec_classoftruncations}, and the
vertex construction in \autoref{subsec_vertex_functions}. The flow
equations for the fully momentum-dependent propagators are derived in
the subsequent subsections. The consistency analysis in
\autoref{sec_lambdas} constrains the momentum-independent parts of the
vertex functions, and is crucial for the global properties
of the phase diagram. The latter is the topic of the first section of
the results, \autoref{sec_phase_diag}. Moreover, the reliability of
all results is tested within a regulator study. To this end we use the
optimized regulator as well as a class of exponential ones. The
results are found to be stable against variations of the
regulator. The properties of the \UV{} regime are discussed in
\autoref{sec_UV_regime}. In this section we also investigate the stability and 
reliability
of the derivative expansion, at the basis of the full
momentum-dependence. In \autoref{sec_IR_regime} the properties of the
\IR{} regime are discussed.  

\section{Flows in quantum gravity}\label{sec_classoftruncations}

A quantum field theory is entirely described by a complete set of correlation
functions. The generating functional for the 1PI correlators is the
effective action $\Gamma[\bar{g},\phi]$, where we have already introduced a 
fixed
background metric $\bar{g}$ and a fluctuation super-field $\phi$. 
In the case of gravity this super-field is given by the vector $\phi =
(h,\bar{c},c)$, where $h$ is the graviton field and $c,\bar c$ are
the corresponding ghost and anti-ghost fields.

In the present work we use the functional renormalization group approach, for 
reviews 
on quantum gravity see
\cite{Litim:2003vp,Niedermaier:2006wt,Percacci:2007sz,Codello:2008vh,
Litim:2011cp,Reuter:2012id,Nagy:2012ef}, for general 
reviews and other applications   
see
e.g.~\cite{Litim:1998nf,Berges:2000ew,Aoki:2000wm,Bagnuls:2000ae,Polonyi:2001se,
Pawlowski:2005xe,%
Gies:2006wv,Schaefer:2006sr,Rosten:2010vm,Pawlowski:2010ht,%
Scherer:2010sv,Braun:2011pp,Metzner:2011cw,Boettcher:2012cm,2012LNP...852...49D}
. 
With the functional renormalization group, the effective action can be
determined via a
functional differential equation for the scale-dependent effective action
$\Gamma_k[\bar{g},\phi]$, which for quantum gravity is given by 
\cite{Reuter:1996cp} 
\begin{align}\label{floweq}
\notag \partial_t \Gamma_k[\bar{g},\phi] = \frac{1}{2} & \mathrm{Tr} \,
\left[\0{1}{\Gamma^{(2h)}_{k} + R_{k,h}} \,\partial_t R_{k,h}
\right][\bar{g},\phi]
\\[2ex]  - & \mathrm{Tr} \, \left[ \0{1}{\Gamma^{(\overline{c}c)}_{k} +
R_{k,c}}\,\partial_t
R_{k,c}\right][\bar{g},\phi] \, .
\end{align}
It involves an \IR{} regulator $R_k$ which is
implemented on the level of the path
integral and carries an \IR{} cutoff-scale $k$.  
In addition to that, $t$ denotes the logarithmic \RG{} scale, $t := \log 
(k/k_0)$,
with an
arbitrary normalization scale $k_0$, and the Tr implies an integral
over all continuous and a sum over all discrete indices. We will also make use
of the notation $\partial_t f(k) =: \dot{f}(k)$ for any scale-dependent
quantity.
Moreover, we have introduced the
notation 
\begin{equation}
\Gamma^{(\phi_1 ... \phi_n)}_k[\bar{g},\phi] \colonequals 
\frac{\delta^n \Gamma_k[\bar{g},\phi]}{\delta \phi_1 \cdots \delta
\phi_n} \, 
\end{equation}
for the 1PI vertex functions, which are derivatives of the effective action with
respect to the fluctuation fields and are the elements of the full super
space matrix $\Gamma^{(n)}_k$. 

The right hand side of the flow equation \eqref{floweq} depends on the
the two-point correlators of the fluctuation field $\phi$. It is
important to note that the fluctuation correlation functions do not
agree with the background correlations, i.e.\
\begin{equation}
\left.\frac{\delta^2 \Gamma_k[\bar{g},\phi]}{\delta h^2}\right|_{\phi=0} \neq
\frac{\delta^2 \Gamma_k[\bar{g},0]}{\delta \bar{g}^2} \, ,
\end{equation}
for details see
\cite{Pawlowski:2002eb,Litim:2002ce,Braun:2007bx,Folkerts:2011jz,%
  Christiansen:2012rx}. In other words, one cannot
extract dynamical couplings from the flow of the background field
effective action at vanishing fluctuation fields, $\phi=0$. This
directly relates to the fact that the flow equation \eqref{floweq} for
the effective action at $\phi=0$ is not closed within the standard
background field approach. More importantly, avoiding unphysical
background contributions can be crucial for capturing the correct
non-perturbative physics. In conclusion, in the \FRG{} setup, this
strongly suggests to start from the exact equation for the inverse
fluctuation propagator.

Moreover, the master flow equation \eqref{floweq} leads to an infinite
hierarchy of coupled partial integro-differential equations for the
scale-dependent vertex functions $\Gamma^{(n)}_k$.  More precisely, the
equation for the $n$-point vertex function contains vertex functions
of order $n+1$ and $n+2$.  This system is usually not exactly
solvable. Therefore, one has to employ certain approximation schemes.
We assume that the effective action can be expanded in a functional
Taylor-series around the fixed background metric $\bar{g}$. Moreover,
we choose the flat Euclidean metric, i.e.\ the identity $\bar{g}_{\mu
  \nu} = \delta_{\mu \nu}$, as the expansion point. We will need this
expansion up to fourth order in the graviton field.  In symbolic
notation, the effective action takes the form
\begin{align}
\notag \Gamma_k[\bar{g},\phi] = & \sum_n \frac{1}{n!}
\Gamma^{(n)}[\bar{g},0]
\phi^n 
\\ \notag = & \Gamma_k[\bar{g},0] +
\Gamma^{(h)}_{k}[\bar{g},0] h +  
\Gamma^{(2h)}_{k}[\bar{g},0] h^2 
\\ \notag & + \Gamma^{(3h)}_{k}[\bar{g},0] h^3 
+ \Gamma^{(4h)}_{k}[\bar{g},0] h^4 +
\ldots
\\  & + \Gamma^{(\overline{c}c)}_{k}[\bar{g},0] \overline{c} \, c
 + \ldots \, .
\label{general_exp}\end{align}
The first and second term are of order $h^0$
and $h^1$ and do not enter the RHS of the flow equations for
any $\Gamma^{(n)}_k$. 

The above expansion of the effective action in powers of the fluctuating field
around a flat Euclidean background $\bar g_{\mu \nu} = \delta_{\mu \nu}$
also restricts the number of higher
derivative operators which can contribute to the vertex functions of $n$-th
order. For instance, the most general form of the two-point function derives
from an action which includes at most $\mathcal{O}(R^2)$ operators. All higher
order terms vanish after two functional differentiations and evaluation on a
flat background. Terms with higher order than this only contribute to vertex
functions of order $n>2$.

\section{Vertex expansion}\label{sec_presenttruncation} 
In the present work we use the systematic vertex expansion scheme as
suggested in \eqref{general_exp}. The hierarchy of flow equations that
has been introduced in the last section has to be truncated at finite
order. This means that one can calculate the flow of a vertex function
of given order $n$ and use an ansatz for $\Gamma^{(n+1)}_k$ and
$\Gamma^{(n+2)}_k$. We will compute the basic quantity of the present
approach, the full two-point correlation functions of the fluctuation
fields $\phi$, that is $n=2$.  The corresponding flows rely on the
two- but also on the three- and four point functions of the
fluctuation fields. Hence we also introduce approximations for
$\Gamma^{(3)}_k$ and $\Gamma^{(4)}_k$ that are consistent with the
symmetries of the theory and have the correct \RG{}-scaling. The latter
property is essential for the global \UV{}--\IR{} flows considered
here.

\subsection{Structure of the vertex
functions}\label{subsec_vertex_functions}

First of all, we have to specify the tensor structures of the vertices. In the
present work, we use the classical tensor structures which arise from functional
differentiation of the Einstein-Hilbert action. The gauge-fixed
Einstein-Hilbert action, including the ghost part, is given by
\begin{align}\label{EH_action}
S = \frac{1}{16 \pi G_N} &\int \text{d}^4x
\sqrt{\text{det}g}\;(-R+2\Lambda) \notag \\
  + &\int\text{d}^4x \sqrt{\text{det}\bar{g}}\; \bar{c}^\mu
\mathcal{M}_{\mu\nu} c^\nu \\  
  +  &\int\text{d}^4x
\sqrt{\text{det}\bar{g}}\; \frac{1}{2 \xi} \bar{g}^{\mu \nu} F_\mu F_\nu \,. 
\notag
\end{align}
In \eqref{EH_action}, $G_N$ is the Newton constant and $\Lambda$ is the
cosmological constant.  The Fadeev-Popov operator is given by
\begin{equation}
\mathcal{M}_{\mu\nu} =\bar{\nabla}^{\alpha} (g_{\mu \nu} \nabla_{\alpha}
+ g_{\alpha \nu} \nabla_{\mu}) -\bar{\nabla}_{\mu}\nabla_{\nu} \, ,
\end{equation}
and the linear gauge fixing conditions reads
\begin{equation}
F_\mu =
\bar{\nabla}^\nu h_{\mu \nu} -\tfrac{1}{2} \bar{\nabla}_\mu h^{\nu}_{~\nu} \,.
\end{equation}
Moreover, in this work we restrict ourselves to Landau gauge, that is
$\xi\rightarrow0$. Extended ghost interactions are studied e.g.\ in
\cite{,Eichhorn:2009ah,Eichhorn:2013ug}.

The standard Einstein-Hilbert truncation amounts to replacing the
gravitational coupling and the cosmological constant in
\eqref{EH_action} by running couplings $G_{N,k}$ and $\Lambda_k$. The
vertex functions are then given by functional derivatives of this
effective action.  However, this approximation turns out to be
inconsistent in the physical \IR{} limit, which will be discussed
in detail later. We also mention that the basic Einstein-Hilbert
truncation does not disentangle the difference between a wave-function
renormalization and a running coupling, since the running of the
latter is simply identified with the running of the former. Note that
in Yang-Mills theory such an approximation gives a deconfining
potential of the order parameter even in the confining regime, see
\cite{Braun:2007bx,Marhauser:2008fz,Fister:2013bh}. In this case, it
is also the non-trivial momentum-dependence of the correlation
functions that plays a crucial r$\hat{\rm o}$le for capturing the
correct non-perturbative physics. Additionally, in the \UV{} limit,
$k\to\infty$, the full momentum-dependence of correlation functions is
potentially relevant. In particular, a derivative expansion implies
$p^2/{\rm scale}^2\ll 1$ which relates to low energy physics. So far,
these momentum-dependencies have not been taken into account.
 
Consequently, we construct more general vertex functions that take
into account the above properties while keeping the classical tensor
structures. The construction of such vertex expansions of the
scale-dependent effective action was introduced in \cite{Fischer:2009tn}
and applied in the context of Yang-Mills theories.  A similar
truncation based on these ideas was recently applied in quantum
gravity to \cite{Codello2013}.  One guiding principle in this
construction is \RG{} invariance, i.e.\ invariance of the full
effective action under a change of the renormalization scale $\mu$:
\begin{equation}
\mu \frac{\mathrm{d}}{\mathrm{d} \mu} \Gamma = 0 \, , 
\end{equation}
where $\mu$ should not be confused with the running \IR{} cutoff
scale $k$, for a detailed discussion see \cite{Pawlowski:2005xe}. In
addition to that, we parameterize the vertex functions, i.e.\ the
coefficients in the expansion \eqref{general_exp} schematically as
\begin{equation}
\Gamma^{(n)} = Z^{\0{n}{2}} \, \bar{\Gamma}^{(n)} \,  
\end{equation}
with a $\mu$-independent part $\bar{\Gamma}^{(n)}$ and a
$\mu$-dependent wave-function renormalization $Z(p^2)$ of the attached
fields. The above construction implies the correct scaling behaviour
for the fields according to
\begin{equation}
\mu \frac{\mathrm{d}}{\mathrm{d} \mu} \phi = \eta\, \phi \, . 
\end{equation}
In the present work we use a uniform wave function renormalization,
$Z_h = Z_{h_i}$, for all components of the graviton. Non-uniform
$Z$-factors will be subject to a forthcoming publication
\cite{Vertices}. The transverse-traceless ($\rm TT$) part of the full
propagator is now parameterized as
\begin{equation}\label{2point}
\Gamma^{(2h)}_{\mathrm{TT}}(p^2) = Z_{h} (p^2) (p^2 - M^2)
\Pi_{\mathrm{TT}}(p) \,,
\end{equation}
with the transverse-traceless projector $\Pi_{\mathrm{TT}}(p)$, and an
effective mass term $M$ representing the momentum-independent part of
the two-point function. Note that the $Z$-factors are functions of the
covariant Laplacian $\Delta$, and therefore include infinitely many
terms in a covariant expansion of the effective action in powers of
the Laplace operator. This is the first \RG{} study of quantum gravity
taking into account this general momentum dependence of the graviton
and the ghost propagator.  The fully momentum-dependent $Z$-factors
lead to a vertex construction with the required \RG{} scaling
properties. They also embody a corresponding implicit, non-canonical
momentum-dependence of the vertex functions, in line with the
co-linear singularity structure of vertex functions. Note that a
similar vertex construction is achieved by simply taking further
$h$-derivatives of the two-point correlation function \eqref{2point} 
with $p^2 \to \Delta(\bar g,h)$. 

Finally, we allow for additional running parameters $\Lambda^{(n)}_k$
which govern the (consistent) scale-dependence of the
momentum-independent part of the vertex functions. This takes into
account scaling properties of the vertex functions that are crucial
for the global flows structure, and has not been considered before.
The construction of the vertex functions also include appropriate
powers of a scale-dependent Newton coupling $G_{N,k}$ as prefactors of
the vertex functions. These factors, apart from the wave function
renormalization factors, encode the correct scale-dependence of the
vertices, see \cite{Fischer:2009tn}. Note that in general there are
separate coupling constants $G^{(n)}_k$ for each vertex function, but
we identify $G^{(n)}_k = G_{N,k}^{\frac{n}{2}-1}$ in the present
work. Still, this construction goes far beyond the approximations
considered so far in \RG{}-gravity. In summary, the vertex functions take
the form
\begin{equation}
\Gamma^{(\phi_1 \dots \phi_n)}_k = \prod\limits_{i=1}^n
\sqrt{Z_{\phi_i,k}(p_i)}\, G_{N,k}^{\frac{n}{2}-1} \,
\mathcal{T}^{(n)}_k(p_1,\dots,p_n;\Lambda^{(n)}_k) \, ,
\label{vertex_functions}\end{equation}
with tensor structures
\begin{equation}
\mathcal{T}^{(n)}_k= S^{(n)}(p_1,\dots,p_n;G_N
= k^2,\Lambda \rightarrow \Lambda^{(n)}_k) \, ,
\label{tensor_struc}\end{equation}
that arise from functional differentiation of the classical
Einstein-Hilbert action $S$ given in \eqref{EH_action}. We left out
the momentum arguments on the LHS as well as the functional dependence
on the fields. If we leave out the functional argument, it is
implicitly understood that the functional is evaluated at vanishing
fluctuation fields and on a flat background.  The tensor structures
$\mathcal{T}^{(n)}_k$ carry not only the canonical explicit
momentum-dependence of the vertex functions, but also the running
parameters $\Lambda^{(n)}_k$. They are defined by
\begin{equation}\label{eq_lambda_n}
  \mathcal{T}^{(n)}_k(p_i=0;\Lambda^{(n)}_k) \equalscolon -2 \Lambda^{(n)}_k
  \tilde{\mathcal{T}}^{(n)}(\delta_{\mu \nu}) \, ,\end{equation}
where $\tilde{\mathcal{T}}^{(n)}(\delta_{\mu \nu})$ is the tensor structure
arising from functional differentiation of the integral of the volume form with
respect to the metric tensor. The factor of $-2$ in the definition is
introduced such that we recover the classical cosmological constant in the
Einstein-Hilbert action. We emphasize again that the consistent \RG{}-scaling of 
the parameters 
$\Lambda^{(n)}_k$ is of major importance for the transition from the \UV{} 
regime  
to the \IR{} regime, as well as for the stability of
the \IR{} regime. In this regime the $\Lambda^{(n)}_k$ are determined via a 
self-consistency
analysis of the scaling behaviour. Details will be presented in the 
corresponding results sections. 

An example for the above vertex construction is the scalar coefficient
of the TT-two point function of the graviton,
\begin{equation}
Z_h(p^2)(p^2 -2
\Lambda_k^{(2)})\, , 
\label{EH_kernel} \end{equation} 
see also \eqref{2point}. The tensor structures for the general two-point 
function, from which the
above results via TT-projection, are given by
\eqref{tensor_struc} and arise from functional differentiation of the
Einstein-Hilbert action. Equation \eqref{EH_kernel} and \eqref{2point} entail 
that 
$\Lambda_k^{(2)}$ is the effective graviton mass, 
\begin{equation}
M^2_k = -2 \Lambda_k^{(2)} \, . 
\label{def_M}\end{equation}
Once again, note that this mass term is not the cosmological constant.

\subsection{Flow of the propagator}\label{propagator_flow}

\begin{figure}
\begin{align*}
 \partial_t \left.
\frac{\delta^2\Gamma[\bar{g},\phi]}{\delta h^2}
\right|_{\phi=0}(p^2) &= -\frac{1}{2}
\includegraphics[height=6ex]{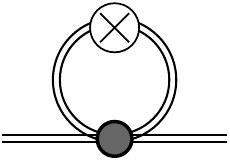} +
\raisebox{-2ex}{\includegraphics[height=6ex]{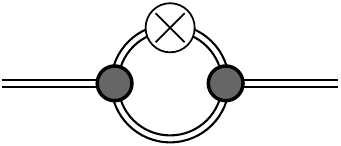}}\\
	      & \mathrel{\hphantom{=}} - 2
\,\raisebox{-2.2ex}{\includegraphics[height=6ex]{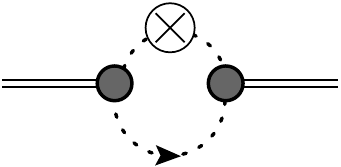}}  
		 \equiv \text{Flow}^{(2h)}(p^2)\\
 \partial_t \left.\frac{\delta^2\Gamma[\bar{g},\phi]}{\delta c \delta \bar{c}}
\right|_{\phi=0}(p^2) &=
\raisebox{-2.0ex}{\includegraphics[height=6ex]{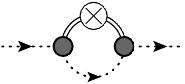}} +
\raisebox{-2.8ex}{\includegraphics[height=6ex]{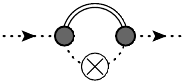}} \\
		 &\equiv \text{Flow}^{(\overline c c)}(p^2)
\end{align*}
\caption{Diagrammatic representation of the flow of the second order vertex
functions. The dressed graviton propagator is represented by a double line,
the
dressed ghost  propagator by a dashed line, while a dressed vertex is denoted
by a dot
and the regulator insertion by a crossed circle.}
\label{fig:flow_of_prop}
\end{figure}

The flow of $\Gamma^{(2h)}$ in a standard Einstein Hilbert truncation
has been calculated in \cite{Christiansen:2012rx}. The corresponding
flow equation is obtained by functional differentiation of
\eqref{floweq}, and its diagrammatic representation is shown in
\autoref{fig:flow_of_prop}.

In the present work, we compute the fully momentum-dependent two-point
functions with the \RG{}-consistent ansatz \eqref{vertex_functions} for
the three- and the four-point function.  With
\autoref{fig:flow_of_prop} and \eqref{vertex_functions} the flows of
the two-point functions $\Gamma^{(2)}$ depend on
\begin{equation}\label{couplings}
\left(Z_h(p^2),Z_c(p^2),M^2, G_N,\Lambda^{(3)},\Lambda^{(4)}\right) \,. 
\end{equation}
Here we have already dropped the subscript $k$, and if
not stated otherwise, all generalized couplings are scale-dependent.  

The two-point function contains all tensor structures in the York
transverse-traceless decomposition. The transverse-traceless part is
not constrained by Slavnov-Taylor identities and is expected to carry
the essential properties of the graviton. In this work we identify all
wave function renormalizations of the graviton modes with that of the
transverse-traceless mode. Hence, from now on, all expressions for the
two-point functions relate to the TT-part. A study with all
independent tensor structures will be presented elsewhere
\cite{Vertices}.

In this work, we use a regulator of the form
$R_{\phi}(p^2) = Z_{\phi}(p^2) R^0_{\phi}(p^2)$ with $R^0_{\phi}(p^2)=
p^2 \, r(p^2) \mathcal{T_{\phi}}$, where $\mathcal{T_\phi}$ denotes
the tensor structure of the corresponding two-point function evaluated
at vanishing mass, and $r(p^2)$ is a dimensionless shape function.
Then, the flow of
the inverse propagator reads
\begin{align}\label{flowofprop2}
\notag \partial_t \Gamma^{(2h)}(p^2) & = (p^2+M^2) \partial_t Z_h(p^2) +
Z_h(p^2) \partial_t M^2 \,
\\ &= \text{Flow}^{(2h)}(p^2) \,,
\end{align}
where 
\begin{align}
&\text{Flow}^{(2h)}(p^2) = G_N Z_h(p^2) \times \label{flow_structure} \\ 
& \int \text{d}^4q 
\sum_{\phi} \left(\partial_t r(q^2) + \frac{\partial_t Z_\phi (q^2)}{Z_\phi
(q^2)} r(q^2) \right) I_\phi (p^2,q^2,\Lambda^{(n)}) \, . \notag
\end{align}
In the above equation, $I_\phi (p^2,q^2,\Lambda^{(n)})$ are scalar
functions that arise from the contraction of the diagrams and a
subsequent projection onto the TT-structure, and $n=2,3,4$.  This
structure follows from our vertex construction discussed above.
Explicit expressions for the flow equations are given in 
Appendix~\ref{app_flow_eq}. For the ghost sector, we apply the same
strategy and arrive at the much simpler equation
\begin{equation}
p^2 \partial_t Z_c(p^2) = \text{Flow}^{(\overline c c)}(p^2) \, .
\label{ghost_flow}\end{equation}
Note that with the \RG{} consistent vertex ansatz \eqref{vertex_functions}
and the structure of the flow equations for $\Gamma^{(n)}$, one can infer that
the wave-function renormalization $Z(p^2)$ does never enter a flow equation
alone, but always in the combination $\dot{Z}/Z$. This
motivates the definition of the anomalous dimensions
of the graviton,
\begin{align}\label{defeta}
 \eta_h(p^2) &:= - \frac{\partial_t Z_h(p^2)}{Z_h(p^2)} \, , \\
\intertext{and of the ghosts,}
\eta_c(p^2) &:= - \frac{\partial_t Z_c(p^2)}{Z_c(p^2)} \, .
\end{align} 
We note that the RHS of the flow of the $\Gamma_k^{(2)}$ does depend
on $\eta_h,\eta_c,M^2$ and the three- and four-point functions, see
\eqref{couplings}. In the standard Einstein-Hilbert setup, a
scale-dependent gravitational coupling is constructed from the
graviton wave-function renormalization, and the momentum-independent
parts of the vertex functions are all identified with the cosmological
constant, $\Lambda^{(n)} \equiv \Lambda$.  This procedure closes the
equations \eqref{flowofprop2} and \eqref{ghost_flow} and was used at
least in parts in all \FRG{} gravity calculations so far. 

In the
present work this identification, which spoils the scaling properties
of the correlation functions, is avoided. For the gravitational
coupling constant $G_N$, we use the relation of the present framework
at flat backgrounds to that with geometrical effective actions, see
\cite{Donkin:2012ud}. This is discussed in more detail below. The
couplings $\Lambda^{(n)}$ are constrained within a self-consistency
analysis, see \autoref{sec_lambdas}.

Finally we introduce dimensionless, scale-dependent
couplings, to wit 
\begin{align}
 g &:= G_N k^2\,, & \mu &:= M^2\, k^{-2},  
\\ \lambda &:= \Lambda k^{-2}\, , & \lambda^{(n)} &:= \Lambda^{(n)} k^{-2}
\,,
\label{dimless}\end{align}
with $n\geq3$ and $\Lambda = \Lambda^{(1)}$. It is left to project the
functional flow onto individual flow equations for all running
couplings. 

\subsubsection{The running mass}
The flow equation for the mass $M^2$ is obtained from the flow of the
inverse propagator, evaluated at the pole of the propagator, i.e.\
$p^2 = -M^2$. Taking the $t$-derivative of the on-shell
two-point function, $\Gamma^{(2h)}(-M^2)$, yields
\begin{equation}
\begin{aligned}
 0 &=  \left. \partial_t \Gamma^{(2h)}(p^2)\right|_{p^2=-M^2} - Z_h(-M^2)
\partial_t
M^2 \, .
\end{aligned}
\end{equation}
Solving for the running of the mass parameter, we get
\begin{equation}
 \partial_t M^2 = \frac{\partial_t \Gamma^{(2h)}(-M^2)}{Z_h(-M^2)} \, .
\label{runningmass}\end{equation}
One of the goals of this work is to evaluate the phase diagram of quantum 
gravity and its
fixed point structure. For this reason, it is convenient to derive $\beta$-
functions for the dimensionless parameters. Then, the above equation 
translates into
\begin{align}
\notag \partial_t \mu &= - 2\mu + \frac{\partial_t \Gamma^{(2h)}(-M^2)}{k^2
Z_h(-M^2)}
\\ &=:\beta_{\mu}[\eta_h,\eta_c](g,\mu,\lambda^{(3)},\lambda^{(4)}) \,
\, .
\label{flow_mu}\end{align}
The explicit form of the $\beta$-function is given in
Appendix~\ref{app_flow_eq}.  It is clear from
\eqref{flow_structure} that the $\beta$-function for the running mass
shows a functional dependence on the anomalous dimensions.  It turns
out that the final result for the $\beta$-function of $\mu$ and the
momentum- dependent equation for the anomalous dimension $\eta(p)$
does not depend on the projection point, see Appendix~\ref{app_anomproj}. The 
finite difference at $p^2 =-M^2$ is just the most
convenient choice.

\subsubsection{Integral equations for the anomalous dimensions}

Starting from the general equation \eqref{flowofprop2}, using the
above definition of the anomalous dimension and inserting \eqref{runningmass},
we obtain an integral equation for $\eta_h$ which reads
\begin{equation}
\eta_h(p^2) = -\frac{\dfrac{\partial_t \Gamma^{(2h)}(p^2)}{Z_h(p^2)} -
\dfrac{\partial_t \Gamma^{(2h)}(-M^2)}{Z_h(-M^2)}}{p^2+M^2} [\eta_h,\eta_c]
\label{etanice} \,. 
\end{equation}
Note that all isolated $Z$-
factors drop out, see \eqref{flowofprop2}. 
The same procedure can be applied for the ghost sector. Since there is no ghost
mass, we trivially arrive from \eqref{ghost_flow} at
\begin{equation}\label{etanicec} 
\eta_c (p^2) = -\frac{\partial_t \Gamma^{(\overline{c} c)}}{p^2
Z_c(p^2)}[\eta_h,\eta_c] \, . 
\end{equation}
The explicit form of eqs.~\eqref{etanice} and \eqref{etanicec} is given in 
Appendix~\ref{app_flow_eq}.
The full expressions of  are derived and solved numerically, with the help
of \textsc{Form} \cite{FORM}, \textsc{xTensor} \cite{xTensor}, and 
\textsc{Eigen} \cite{EigenWeb}. 

\subsection{The running gravitational coupling
\texorpdfstring{$g$}{g}}\label{sec_gdot}

As already mentioned, we use geometrical flow equations for $G_N$ in
order to close the system of differential equations. This approach
allows an inherently diffeomorphism- invariant construction of flows
in quantum gravity, see \cite{Branchina:2003ek,Pawlowski:2003sk}. It
has been applied to the phase structure of quantum gravity in
\cite{Donkin:2012ud}, where evolution equations $\beta_g$ for the
dynamical coupling $g$, and $\beta_{\bar{g}}$ for the background
coupling $\bar g$ are derived. In the present work we utilize the fact that the
geometrical approach is directly related to the present approach in a
flat background. In particular, the dynamical and background couplings
in both approaches agree.

Moreover, with the fully momentum-dependent anomalous dimensions
computed in the present work, we are able to directly incorporate
effects of arbitrarily high powers of derivatives in the equations for
$\beta_g$,$\beta_{\bar{g}}$ in \cite{Donkin:2012ud}.  The anomalous
dimensions enter the geometric flow equations in very much the same
way as in \eqref{flowofprop2}. However, the
wave-function renormalizations in \cite{Donkin:2012ud} are
momentum-independent and can be pulled outside the integrals. This is
not the case in our set-up. Entering the equations with the
momentum-dependent $\eta_\phi(p^2)$ calculated via \eqref{etanice}
leads to a modification on the level of the threshold functions
$\Phi$. These modified threshold functions are given in 
Appendix~\ref{app_flow_eq}.

With these ingredients, the 
general structure of the $\beta$-function for the dynamical gravitational
coupling is given by
\begin{equation}
\beta_{g}[\eta_h,\eta_c]\left(g,\mu \right) = 2g +
F_g[\eta_h,\eta_c]\left(g,\mu
\right) \, ,
\end{equation}
and the one for the background coupling takes the same form with $g$
being replaced by $\bar{g}$ and an individual loop contribution
$F_{\bar{g}}[\eta_h,\eta_c]\left(\bar{g},\mu \right)$. The functionals
$F_g$ and $F_{\bar{g}}$ are given in Appendix~\ref{app_flow_eq}.  Note
that the flow equation of the background coupling depends on the
dynamical coupling via the anomalous dimensions, while the converse
does not hold.

\subsection{The couplings 
\texorpdfstring{$\Lambda^{(n)}$}{Lambdan}}
\label{sec_lambdas}
In \autoref{subsec_vertex_functions} we have introduced an
approximation which takes into account scale-dependent couplings
$\Lambda^{(n)}$ for the momentum-independent part of each vertex function. For
the second order, we have identified $\Lambda^{(2)}$ as the
graviton mass $M^2$, see \eqref{def_M}. In the present section we discuss the 
vertices with $n\geq3$.

The Einstein-Hilbert truncation, which identifies all $\Lambda^{(n)}$
with the cosmological constant $\Lambda$, is ill-defined in the limit
$\mu\rightarrow -1$. As we will see in \autoref{sec_results}, this
limit is approached by physical \RG{} trajectories in the deep \IR{}.
This regime is crucial to understand the global phase structure of
Euclidean quantum gravity: the couplings $\Lambda^{(n)}$ play a
distinguished role, as the related singularities arise from
the momentum-independent parts of the vertex functions. In order to
cure the inconsistencies of the Einstein-Hilbert truncation, we deduce 
the singularity structure of the couplings $\Lambda^{(n)}$ with $n \geq 3$. The 
full details
are given in Appendix~\ref{appendix_analysis}.  Essentially, the idea
is to expand the right-hand sides of the flow equations for the
$n$-point functions in powers of $1+\mu$ and taking into account the
singularities of highest order. Thus, for $\mu\to -1_+$ we use the
ansatz
\begin{equation}
\lim_{\mu\to -1_+}\lambda^{(n)} \sim  (1+\mu)^{\alpha_n} \, ,
\end{equation}
for $n \geq 3$.
We proceed by inserting this ansatz in the flow equations for $\Gamma^{(nh)}$
and analyze the generic loop integrals to leading order in the singularities
that arise in the limits under consideration.
Consistent scaling of both sides of the flow equations
for arbitrary $n$ leads to the relations
\begin{equation}
\alpha_n = \alpha_{n-2} + \alpha_4 -1 \,,
\end{equation}
for $n \geq 5$ and
\begin{equation}
\alpha_4 \leq 2 \alpha_3 -1 \, . 
\label{final_inequality0}\end{equation}
The parameter $\alpha_4$ obeys the bound  
\begin{equation}
\alpha_4 < 0 \,.  
\end{equation}
The value of the parameters $\alpha_3$ and $\alpha_4$ cannot be
obtained from the divergence analysis alone. They are dynamically
determined by the flow of the three- and four-point function. This
highlights again that the standard Einstein-Hilbert approximation with
$\lambda^{(n)} = - \mu/2$ is inconsistent in the \IR{}, and the
non-existence of the \IR{} fixed point cannot be inferred from such an 
approximation. It is also important to stress that the
qualitative features of the phase diagram do not depend on the
specific choice of $\alpha_3$ and $\alpha_4$, see \autoref{sec_phase_diag} and 
Appendix~\ref{appendix_estimate}. In turn, the
quantitative behaviour does only mildly depend on variations of these
two parameters. Their flows will be studied in a forthcoming publication 
\cite{Vertices}. 

Still, we can estimate $\alpha_3$ based on the saturation of the
inequality \eqref{final_inequality0}. Moreover, the constant parts of
the vertex functions are parametrically suppressed far away from the
singular regime.  This entails that there it is viable to identify
$\Lambda^{(n)}=\Lambda^{(2)}$ as done in all other approximations used
in the literature. From these conditions one obtains
$\alpha_3 \approx -1/9$. More details are given in 
Appendix~\ref{appendix_estimate}. In Appendix~\ref{app_funcform} it is shown 
that
\begin{equation}\label{c_n_param}
 \lambda^{(n)} = -\frac{\mu}{2} (1+\delta\lambda^{(n)})
\end{equation}
is consistent with all constraints, where $\delta\lambda^{(n)}$
parametrizes the deviation from the Einstein-Hilbert approximation. The latter
is modeled by
\begin{equation}\label{Lamodel}
 \delta\lambda^{(n)} = \text{sgn}(\mu) \, \chi \left| \frac{\mu}{1+\mu}
\right|^{-\alpha_n} \, ,
\end{equation}
with $\chi$ a parameter to be tuned to match the aforementioned conditions.

\subsection{The cosmological constant}
It is left to discuss the role of the cosmological constant $\Lambda$
in the present construction. Written on the right hand side of the
field equations, it can be interpreted as an additional source for
gravity. In the classical limit, the quantum equations 
\begin{equation}
\frac{\delta
\Gamma}{\delta \phi} = J_{\mathrm{ext}} \, ,
\end{equation}
with an external source $J_{\mathrm{ext}}$, reduce to the classical
equations of motion. Hence, it is natural to define the cosmological
constant from the one-point function, i.e.\ we identify $\Lambda^{(1)}
= \Lambda$ as the vacuum energy. More precisely, with the vertex
construction \eqref{vertex_functions}, the one-point function takes
the form
\begin{equation}
\left. \frac{\delta}{\delta h} \Gamma \right|_{g=\delta} \sim
 \frac{\Lambda}{\sqrt{G_N}} \sqrt{Z_h}  \, .
\end{equation}
Note that the one-point function does not enter the flow of higher order
vertex functions. Consequently, the cosmological constant decouples from the
$\beta$-functions for Newtons constant, the effective mass and the set of
integral
equations for the anomalous dimensions. On the other hand, these quantities
obviously determine the running of the cosmological constant, i.\,e.\ the 
$\beta$-
function for the dimensionless cosmological constant $\lambda := \Lambda /
k^2$ is of the form
\begin{equation}
\begin{aligned}
\dot{\lambda} & = \beta_{\lambda}[\eta_h,\eta_c](g,\lambda,\mu)
\\& = -2\lambda + g \left( A[\eta_h,\eta_c](\mu) + \lambda\, 
B[\eta_h,\eta_c](\mu)
\right) \,.
\end{aligned}
\end{equation}
The explicit form of this flow equation is given in Appendix~\ref{app_flow_eq}.

\subsection{Regulators and stability}
In order to test the quality of our truncation, we will use several regulators
and vary the parameters $\chi$ and $\alpha_n$ introduced before. As regulators,
on the one hand we use the class of exponential regulators given by
\begin{equation}
 r_a(x) = \frac{1}{x(2e^{x^a}-1)} \, ,
\end{equation}
where $x=p^2/k^2$ is the dimensionless squared momentum.  In our
analysis, we scanned the parameter range $a=\{ 2,3,4,5,6 \}$. On the
other hand, the Litim regulator, \cite{Litim:2000ci}, is used,
\begin{equation}
 r_{opt}(x) = \left( \frac{1}{x} - 1 \right) \theta(1-x) \, ,
\end{equation}
where $\theta(x)$ is the Heaviside step function. Note that this
regulator is optimized within the leading order derivative expansion
but not beyond, see \cite{Litim:2000ci,Pawlowski:2005xe}. Also, 
with the semi-optimized regulator the divergence analysis for the
$\Lambda^{(n)}$ is slightly different from the one performed in 
Appendix~\ref{appendix_analysis},
 but leads to similar results.

We also have scanned different values for the parameters $\chi$ and
$\alpha_3$ in \eqref{Lamodel}, and we have restricted our
investigation to the case of equality in \eqref{final_inequality0}. It
turns out that the results do not depend on the specific choice of
$\alpha_3$. Note that the parameter $\chi$ is bounded from above as
otherwise the parametric suppression of the
$\delta\lambda$-contribution away from the singularity is lifted and
the \UV{} regime is changed. In
\autoref{tab:UVFPparamscan} in Appendix~\ref{appendix_estimate}, a
table is given where the change of the \UV{} fixed point values under
a change of $\alpha_3$ and $\chi$ can be ascertained.

\section{Results}\label{sec_results}
In this section we present our results.  First, the global phase
diagram is discussed. Subsequently, its \UV{} and \IR{} properties
will be examined in more detail. In doing so, we will also make contact
with older results. If not stated otherwise, all results and pictures
are obtained with the specific choice of the exponential regulator
$r_4$.

\subsection{The phase diagram}\label{sec_phase_diag}
\begin{figure}
\includegraphics[width=\columnwidth]{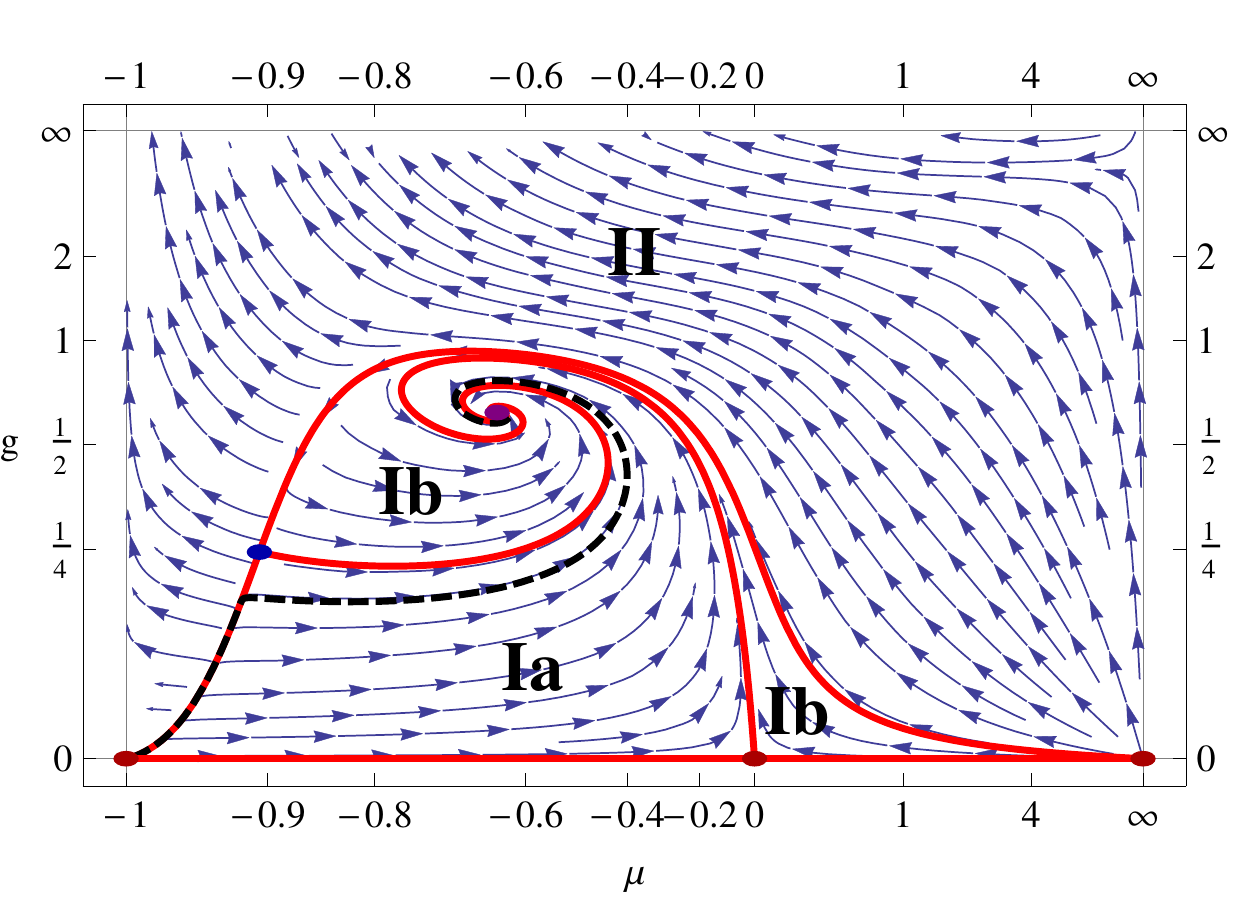}
\caption{Fixed points and global phase diagram in the $(g,\mu)$-plane. Arrows
point from the \IR{} to the \UV{}, red solid lines mark separatrices while dots
indicate fixed points. The black dashed line is a specific trajectory that
connects the \UV{} fixed point with the non-trivial \IR{} fixed point, which is
analyzed further in the text. In analogy to \cite{Christiansen:2012rx}, region 
Ia
corresponds to trajectories leading to the massless IR fixed point, whereas
region Ib leads to the massive IR fixed point. Region II is not connected to the
UV fixed point, and thus physically irrelevant.}
\label{fig:phasediag2}
\end{figure}

The phase diagram for the dynamical couplings $(g,\mu)$ is depicted in
\autoref{fig:phasediag2}.
\begin{figure*}[htb!]
\includegraphics[width=2\columnwidth]{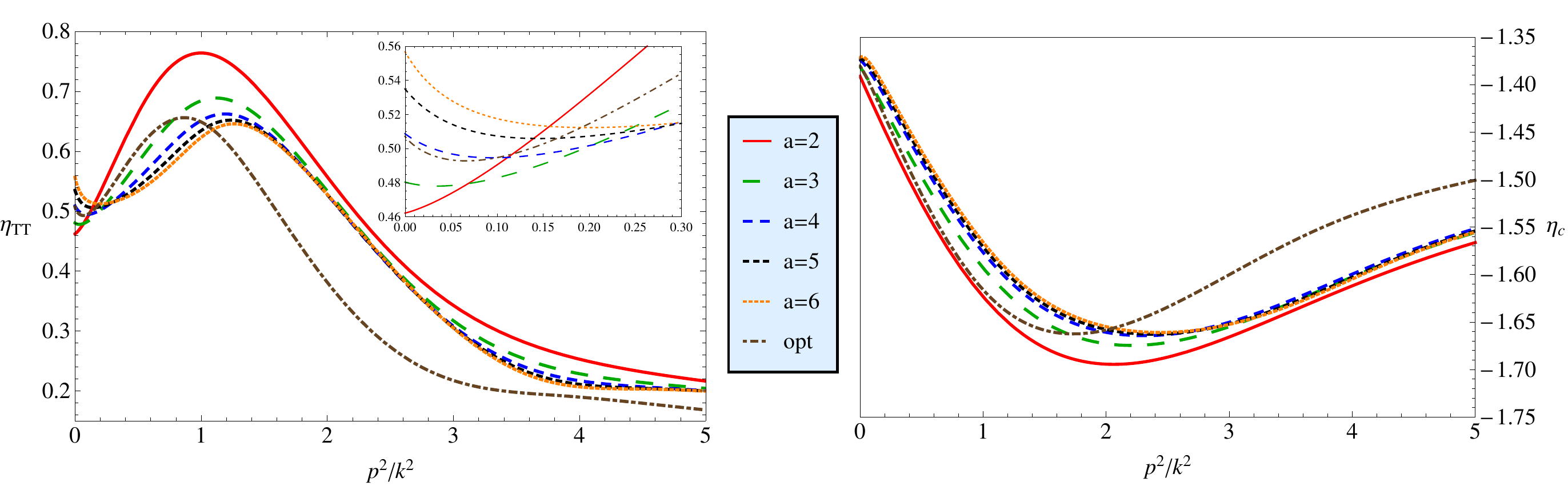}
\caption{The momentum-dependence of the anomalous dimensions of the graviton
(left) and the ghost field (right) for different regulators at the respective
\UV{} fixed points. Only a weak dependence on the parameter is observed. The
difference between optimized and exponential regulators is due to the fact
that the modes are not integrated out at the same scale. From the fact that the
quadratic external momentum terms cancel in the flow of the graviton
\cite{Christiansen:2012rx}, $\eta_h$ goes to zero in the limit of large external 
momenta.
The same is not true for the ghosts, where the anomalous dimension goes to a
constant.}
\label{fig:momdep_eta}
\end{figure*}
We find an attractive \UV{} fixed point with coordinates
\begin{equation}
(g^\text{UV}_*,\mu^\text{UV}_*) = (0.614,-0.645) \, , 
\end{equation}
and complex critical exponents $\theta_{1,2} = (-1.268 \pm 3.009
\mathbf{i})$. This provides further non-trivial evidence for the
asymptotically safe \UV{} structure of quantum gravity. We also 
find the built-in repulsive Gaussian fixed point
at $(g^\text{Gauss}_*,\mu^\text{Gauss}_*) = (0,0)$, and a massive \IR{}
fixed point at $(g^\text{IR}_*,\mu^\text{IR}_*) = (0,\infty)$. The
most striking feature of the present phase diagram is the
confirmation of the attractive massless \IR{} fixed point
\begin{equation}\label{IR}
(g^\text{IR}_*,\mu^\text{IR}_*) = (0,-1) \, ,
\end{equation}
which was already found in \cite{Christiansen:2012rx}, where it
corresponds to a de-Sitter fixed point with $\lambda = 1/2$. This
fixed point implies the global existence of trajectories connecting
the \UV{} fixed point with a finite \IR{} fixed point. The
present result is a clear confirmation that this \IR{} fixed point is
not a truncation artifact, but rather a physical property of the
theory. 

Importantly, it turns out to be an \IR{} fixed point describing
classical gravity. Physical initial conditions lead to globally
defined trajectories that connect the non-trivial \UV{} fixed
point with the physical \IR{} fixed point
$(g^\text{IR}_*,\mu^\text{IR}_*) =(0,-1)$. 

Note also that all \UV{}-complete trajectories are also
\IR{}-complete, and end in either the massive or massless \IR{} fixed
point. In addition to this structure, there is a repulsive fixed point
at $(g^\text{rep}_*,\mu^\text{rep}_*) = (0.250,-0.905)$. This fixed
point was also found in \cite{Donkin:2012ud}.  All essential features
do not depend on the choice of the regulator $r(x)$, and there are
only minor quantitative changes induced by variations of the
latter. The variation of the \UV{} fixed point values under a variation of the 
vertex model parameters $\chi,\alpha_3$, \eqref{Lamodel} is given in 
\autoref{tab:UVFPparamscan} in Appendix~\ref{appendix_analysis}. 

\subsection{\UV{} regime}\label{sec_UV_regime}
\begin{table}
\begin{center}
\begin{tabular}{|c||c|c|c|c|c||c|} \hline
 $a$ & 2 & 3 & 4 & 5 & 6 & opt \\ \hline \hline
 $\mu_*$ & -0.637 & -0.641 & -0.645 & -0.649 & -0.651 & -0.489 \\
\hline
 $g_*$ & 0.621 & 0.622 & 0.614 & 0.606 & 0.600 & 0.831 \\ \hline
 $\overline g_*$ & 0.574 & 0.573 & 0.567 & 0.559 & 0.553 & 0.763 \\
\hline
 $\lambda_*$ & 0.319 & 0.316 & 0.316 & 0.318 & 0.319 & 0.248 \\ \hline
 EVs & -1.284 & -1.284 & -1.268 & -1.255 & -1.244 & -1.876 \\
  & $\pm$3.247\textbf{i} & $\pm$3.076\textbf{i} & $\pm$3.009\textbf{i} &
$\pm$2.986\textbf{i} & $\pm$2.974\textbf{i} & $\pm$2.971\textbf{i} \\
\cline{2-7}
  & -2 & -2 & -2 & -2 & -2 & -2 \\ \cline{2-7}
  & -1.358 & -1.360 & -1.360 & -1.358 & -1.356 & -1.370 \\ \hline
\end{tabular}
\caption{\UV{} fixed point values and eigenvalues for different regulator
parameters $a$, and the optimized regulator, with parameter values
$\alpha_3=-0.1$ and $\chi=0.1$.}
\label{tab:UVFPregscan}
\end{center}
\end{table}
\begin{table}
\begin{center}
\begin{tabular}{|c||c|c|c|c|c|c|c|} \hline
  & here & \cite{Litim:2003vp} & \cite{Christiansen:2012rx} & 
\cite{Donkin:2012ud} &
\cite{Manrique:2010am} & \cite{Codello2013} & \autoref{tab:etader_standard} \\
\hline
 $\overline g_*$ & 0.763 & 1.178 & 2.03 & 0.966 & 1.055 & 1.617 & 1.684 \\
\hline
 $\lambda_*$ & 0.248 & 0.250 & 0.22 & 0.132 & 0.222 & -0.062 & -0.035 \\ \hline
 $\overline g_* \lambda_*$ & 0.189 & 0.295 & 0.45 & 0.128 & 0.234 & -0.100 &
-0.059 \\
\hline
\end{tabular}
\caption{Comparison of the \UV{} fixed point coordinates with earlier results
for the optimized cutoff. Parameter values are $\alpha_3=-0.1$ and $\chi=0.1$.
Methods of the references (in order): background approximation
\cite{Litim:2003vp}, bi-local
projection \cite{Christiansen:2012rx}, geometric approach \cite{Donkin:2012ud},
bi-metric approach \cite{Manrique:2010am}. The mixed approach is applied in
\cite{Codello2013} and is also discussed in the present paper in the last
paragraph of this subsection, \autoref{tab:etader_standard}.}
\label{tab:UVFPcomparison}
\end{center}
\end{table}
Let us further investigate the properties of the \UV{} fixed
point. First of all, the existence of the fixed point does not depend
on the specific choice of the regulator. Moreover, it is attractive in
all four directions investigated here. Furthermore, even though the
critical exponents of the dynamical quantities $(g,\mu)$ are complex,
the ones of the physical background couplings $(\overline g, \lambda)$
are real. This was also found in \cite{Christiansen:2012rx} and
\cite{Codello2013}. Notice
that the eigenvalue corresponding to $\overline g$ is exactly -2,
which can be immediately inferred from the specific structure of the
background coupling flow equation. Also, the eigenvalue corresponding
to $\lambda$ is inherently real, as its flow equation is a polynomial
of order one in the cosmological constant. All these points are
summarized in \autoref{tab:UVFPregscan}. 

The connection to earlier results is drawn in
\autoref{tab:UVFPcomparison}. The present results support the
qualitative reliability of the Einstein-Hilbert type approximations in
the \UV{} regime. 

The couplings as functions of the \RG{} scale $k$
along one selected trajectory (marked as a dashed black line in the phase
diagram)
are shown in \autoref{fig:traj}. One can see how the couplings tend to
their finite fixed point values in the \UV{}. The \IR{} regime
will be discussed below.

\begin{figure*}[htb!]
\includegraphics[width=2\columnwidth]{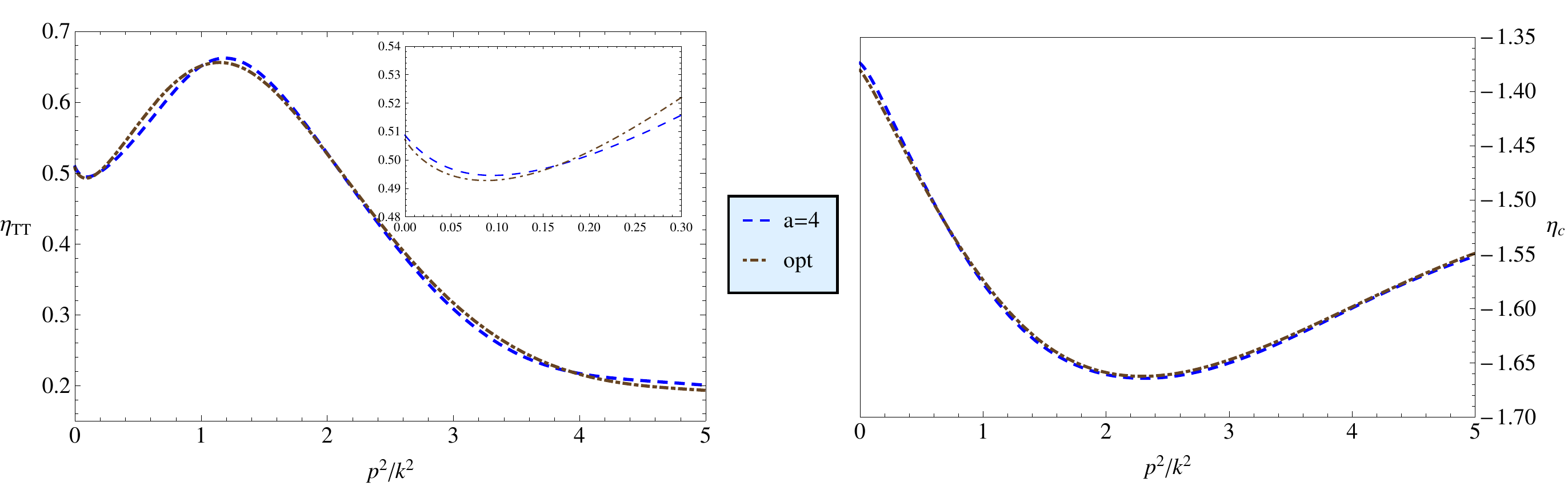}
\caption{Comparison of the momentum dependence of the graviton and the
ghost anomalous dimension with the exponential regulator with $a=4$ on the one
hand, and the optimized regulator with the cutoff-rescaling
\eqref{cutoff_rescaling} on the other.}
\label{fig:reg_rescale}
\end{figure*}

A further quantity of interest is the anomalous dimension. The
momentum-dependence of both graviton and ghost anomalous dimension is given in
\autoref{fig:momdep_eta} for all used regulators at their respective \UV{} fixed
point. As one can see, only quantitative differences occur. The graviton
anomalous dimension is of the order of $0.5$, whereas the ghost anomalous
dimension is of the order of $-1.5$. The difference
between exponential and optimized regulators is due to the fact that the
regulators integrate out modes at different scales. Consequently, the effective 
cut-off
scale is regulator-dependent. A formal discussion of scale-optimization
can be found in \cite{Pawlowski:2005xe}, and is applied in the context of finite
temperature Yang-Mills theory in \cite{Fister:2011uw}. For instance, for the
exponential cutoff $r_a(x)$ with $a=4$, we find that if one
rescales
\begin{equation}
k_{\mathrm{opt}} \to 1.15 \, k_{\mathrm{opt}} \, , 
\label{cutoff_rescaling}\end{equation}
the momentum-dependence of the anomalous dimension with a optimized regulator
matches the one
obtained with an exponential regulator, see \autoref{fig:reg_rescale}.
In general, we observed that the (TT-part of the) graviton anomalous
dimension is positive, however there are indications that this does
not remain so when the other degrees of freedom of the graviton
receive an individual anomalous dimension \cite{Vertices}.  On the
other hand, the ghost anomalous dimension is strictly negative, as was
already found in \cite{Eichhorn:2010tb} and \cite{Groh:2010ta}.  We
also note that the anomalous dimensions are not the leading
contribution to the flow. This means that by setting $\eta(q^2) =0$ on
the RHS of the flow \eqref{flow_structure}, one captures all
qualitative properties dicussed here. Hence, the anomalous dimensions
only constitute correction effects while the leading term on the RHS
of the flow equations is the one proportional to $\dot{r}$. In the
ghost sector this pattern is even more pronounced, and dynamical ghost
effects on the phase diagram and the running couplings are very small. \\[-.5ex]

\paragraph*{Derivative expansion:}

We close this section with a discussion of the stability of the
(covariant) derivative expansion which is the standard approximation
scheme used so far. The first calculation of the graviton anomalous
dimension has been presented in \cite{Christiansen:2012rx} within the
\FRG. There, the flow is projected at $p=k$. In the work
\cite{Codello2013} a derivative expansion around $p=0$ is performed.
The full results in the present study show a strong momentum
dependence of the correlation functions as well as their flows in the
cut-off regime with $p^2/k^2 \lesssim 1$. Such a strong momentum
dependence of the flows either requires higher orders in the
derivatives or a non-local expansion that works-in the information of
momenta close to zero and those close to $p^2/k^2 =1$, see 
\cite{Fister:2011uw,Christiansen:2012rx,Schnoerr:2013bk}. 

Note that there is a strong cut-off dependence of the graviton
anomalous dimension at the \UV{} fixed point for small momenta
$p^2/k^2 \lesssim 0.05$, see \autoref{fig:momdep_eta}. The occurance
of this regime is presumably related to the mass-scale set by the
fixed point value of $\mu$. Here we investigate its impact on the
value of $\eta_{h}$ in the leading order of the derivative
expansion. We also consider a variation of the expansion point. We
also use the present results with full momentum-dependence in order to
investigate the reliability of the derivative expansion. There, the
computation of the anomalous dimension requires
\begin{equation}\label{eq:expandG}
\0{\partial_{p^2} \dot \Gamma_k^{(2h)}}{Z}= -\eta+ \0{\dot Z'}{Z}(x+\mu) + 
\0{Z'}{Z}(2 \mu +\dot \mu) \,, 
\end{equation}  
e.g.\ at vanishing momentum, $x=p^2/k^2=0$. On the other hand, the momentum
derivative 
of $\eta$ gives the relation 
\begin{equation}\label{eq:expandeta}
\0{\dot Z'}{Z} = -\eta' -\0{Z'}{Z}  \eta  \,.  
\end{equation}  
Inserting \eq{eq:expandeta} in \eq{eq:expandG} leads to 
\begin{equation}\label{eq:expandG2}
\0{\partial_{p^2} \dot \Gamma_k^{(2h)}}{Z}=    -\eta -\eta'(x+ \mu)    +
\0{Z'}{Z}\Bigl[(2-\eta) \mu  -\eta\, x +\dot \mu\Bigr] \,.
\end{equation} 
In the lowest order derivative expansion, that is $\Gamma_k^{(2h)} = Z_k(p^2 
+m^2)$, the 
anomalous dimension $\eta_{\rm der}$ is given by (minus) \eq{eq:expandG2}
evaluated at $x=0$. Moreover, the lowest order implies $Z'=0$ and we simply
arrive at 
\begin{equation}\label{eq:lowestorderder}
\eta_{\rm der}=  \eta(0). 
\end{equation} 
For the regulators used in the present work this leads to 
anomalous dimensions 
listed in \autoref{tab:etader}. 
\begin{table}
\begin{center}
\begin{tabular}{|c||c|c|c|c|c||c|} \hline
 $a$ & 2 & 3 & 4 & 5 & 6 & opt \\ \hline \hline
 $\eta_{\rm der}$ & 0.46 & 0.48 & 0.51 & 0.54 & 0.56 & 0.51 \\ \hline
\end{tabular}
\caption{Anomalous dimension $\eta_{\rm der}$ in the lowest order
  derivative expansion derived from the full flow.}
\label{tab:etader}
\end{center}
\end{table}
\begin{table}
\begin{center}
\begin{tabular}{|c||c|c|c|c|c||c|} \hline
 $a$ & 2 & 3 & 4 & 5 & 6 & opt \\ \hline \hline
 $\eta_{\rm der}$ & 0.44 & 0.50 & 0.58 & 0.66 & 0.74 & 0.61 \\ \hline
\end{tabular}
\caption{Full anomalous dimension $\eta_{\rm der}$ in the lowest order
  derivative expansion derived from the full flow.}
\label{tab:etaderfull}
\end{center}
\end{table}
However, the full lowest order derivative expansion takes into account the
$Z'$-terms on the right hand side.
At vanishing momentum there is the relation 
\begin{equation}\label{eq:primelogZ}
\left.\0{Z'}{Z}\right|_{x=0}=\left.\012 \eta'\right|_{x=0}\,.
\end{equation} 
This is easily derived from 
\begin{equation}\label{eq:Z}
Z_k(p^2)=Z_{k_0}(p^2) \exp\left\{-\int_{k_0}^k \0{d\bar k}{\bar k} \eta^{\ 
}_{\bar k}(p^2)\right\}\,,
\end{equation} 
where both $k$ and $k_0$ are in the scaling regime. The latter
condition implies that $Z'/Z=Z'_{\bar k}/Z_{\bar k}(0)$ and
$\eta'=\eta_{\bar k}'(0)$ are independent of $\bar k\in [k_0\,,\,
k]$. Then we conclude that at $x=0$ we have
\begin{equation}\label{eq:primelogZ1}
\0{Z'_k}{Z_k}=\0{Z'}{Z} \0{k^2}{k_0^2}+ \012 \eta'\,\left(1-
\0{k^2}{k_0^2}\right)\,, 
\end{equation} 
for all $k, k_0$ in the scaling regime and we are led to
\eq{eq:primelogZ}. Hence, in the scaling regime (with $\dot \mu=0$)
the full anomalous dimension in the derivative expansion at $x=0$ is
given by
\begin{equation}\label{eq:lowestorderderfull}
\eta_{\rm der}=   \eta\,\left(1 +\012\,\eta'\,\mu \right) \,, 
\end{equation} 
leading to \autoref{tab:etaderfull}. These results seem to be much more
stable then the approximation \eqref{eq:lowestorderder}. 
\begin{table}
\begin{center}
\begin{tabular}{|c||c|c|c|c|c|c|} \hline
 $\alpha$ & 0 & 0.25 & 0.5 & 0.75 & 1 & 1.15 \\ \hline \hline
 $\eta_{\rm der}$ & 0.57 & 0.50 & 0.39 & 0.25 & 0.074 & -0.016 \\ \hline
\end{tabular}
\caption{Anomalous dimension $\eta_{\rm der}$ in the standard derivative
expansion with optimized regulator in an expansion around $p=\alpha k$ with
$(\mu=0,g=1)$.}
\label{tab:etader_alpha}
\end{center}
\end{table}

To complete the present reliability analysis of Taylor expansions in
momenta $p^2$, we also investigate expansions about a general
expansion point $p = \alpha k$. We present results for the optimized
regulator and evaluate the anomalous dimension for
$g=1,\mu=0,\eta_{\mathrm{c}}=0$. The conclusions of this study do
not depend on the choice of these parameters.  As one can see, the
anomalous dimension of the graviton in a derivative expansion strongly
depends on the specification parameter $\alpha$.  This relates to the
fact that such an expansion only works well if the full flow of the propagator
shows a mild momentum dependence. This is not the case for the flow of
the graviton two-point function, see \cite{Christiansen:2012rx}. In
general, even the sign of the
anomalous dimension depends on the specification parameter. We
conclude that a derivative expansion in quantum gravity with
$\alpha=0$ has to be used with great caution. \\[-.5ex]

\paragraph*{Effect of identifications of couplings in the \UV:}

As already mentioned, the present approximation is the first work that
employs individual running couplings for the momentum-independent part
of each vertex function. In particular, the graviton mass term should
not be identified with the cosmological constant. Still, we have
shown, that the full expansion with momentum-dependent wave function
renormalizations and a mass term for the fluctuating graviton $h$
provides \UV{} fixed point results in qualitative agreement with that
of the standard background field approach, if we identify the mass a
posteriori with (minus 1/2 of) the cosmological constant. Within such
an identification we have a deSitter fixed point.

For completeness, we also have investigated a mixed approach: We use a
flat anomalous dimension $\eta_h^{\ }$ in a derivative expansion about
vanishing momentum, or a momentum-dependent one, $\eta_h(p^2)$, for
the fluctuating graviton. In turn, the flows of the graviton mass and
the Newton coupling $g$ are extracted from the flow of the
cosmological constant and the Newton coupling in the background field
approximation. This can be interpreted as an intermediate step towards
the full approximation studied here. Interestingly this leads to a
very small and negative fixed point value for the cosmological
constant, see also \cite{Codello2013} for such a mixed expansion with
a flat anomalous dimension. Our fixed point results for the case with
a flat anomalous dimension are given in \autoref{tab:etader_standard}. They 
are in qualitative agreement with the results of
\cite{Codello2013}. Notably, the results in the mixed approach deviate
from both, the background field results and that of the full
approximation introduced in the present work. We have also checked
that this originates in the identification of the mass term with the
cosmological constant, the given alternative choices for the flow of
the Newton constant do not alter this result.
\begin{table}
\begin{center}
\begin{tabular}{|c||c|c|c|c|c||c|} \hline
 $a$ & 2 & 3 & 4 & 5 & 6 & opt \\ \hline \hline
 $g_*$ & 1.68 & 1.72 & 1.75 & 1.77 & 1.80 & 1.68 \\ \hline
 $\lambda_*$ & -0.064 & -0.076 & -0.088 & -0.100 & -0.110 & -0.035 \\ \hline
 $\eta^{\rm der}_h$ & 0.81 & 0.93 & 1.03 & 1.14 & 1.24 & 0.86\\ \hline
 $\eta^{\rm der}_c$ & -1.08 & -1.05 & -1.04 & -1.03 & -1.01 & -0.75 \\ \hline
\end{tabular}
\caption{Fixed point values $g_*$ and $\lambda_*$ and anomalous dimension
$\eta_{\rm der}$ at this fixed point in the mixed approach.}
\label{tab:etader_standard}
\end{center}
\end{table}

\subsection{\IR{} regime}\label{sec_IR_regime}

\begin{figure}
\includegraphics[width=\columnwidth]{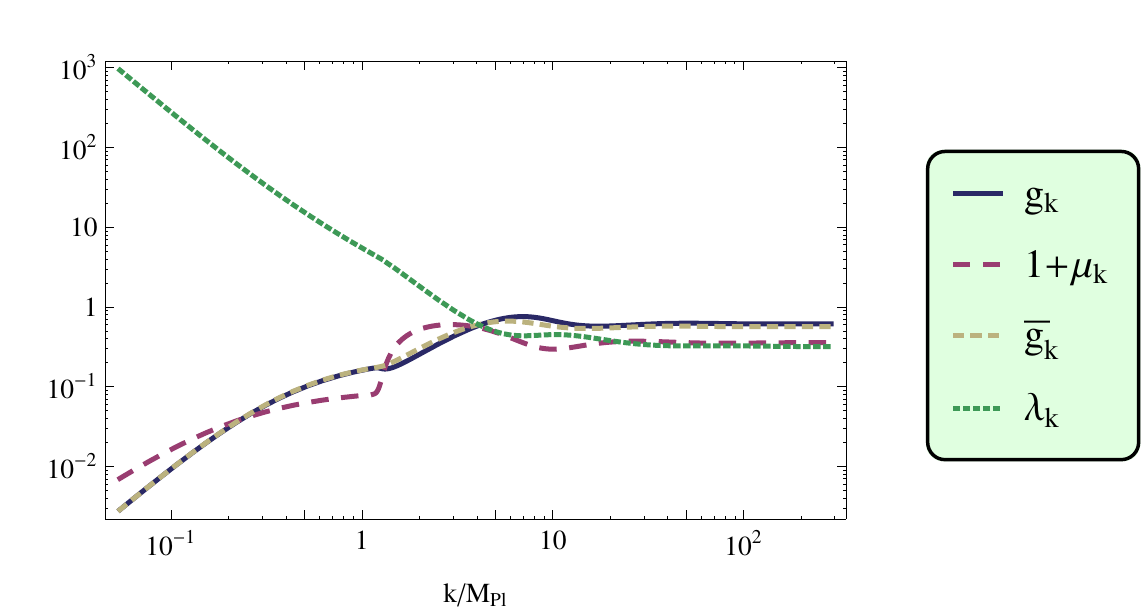}
\caption{The running couplings as functions of the renormalization scale $k$ in
units of the Planck mass. In the \UV{} the couplings tend to their finite
fixed point values. As one follows the trajectories down to the \IR{}, one
can see the scaling behaviour in the vicinity of the \IR{} fixed point.}
\label{fig:traj}
\end{figure}
The non-trivial \IR{} fixed point is located at $(g,\mu)=(0,-1)$. The
most important feature is that it is a classical one, i.e. the
essential couplings scale classically and all quantum contributions
vanish: The gravitational couplings and the cosmological constant scale as
\begin{align}
 g, \overline g \sim& k^2 \, ,& \lambda \sim& k^{-2} \, ,
\label{cl_scaling_g}
\intertext{and the anomalous dimensions vanish,}
\eta_h \to& 0\,, &  \eta_c\to& 0\,, 
\end{align}
see Appendix~\ref{appendix_etaIR}.  This leads to flow trajectories
that connect the asymptotically safe \UV{} regime for
$k\to\infty$ and short distances, with a classical \IR{} regime for
$k\to 0$ and large distances. Since $\mu$ approaches a finite value in
the limit $k \rightarrow 0$, the dimensionful mass $M^2 = \mu \, k^2$
vanishes in the deep \IR{}. Accordingly, this fixed point corresponds
to a massless theory on large distances, which is consistent with
gravity as a force with infinite range.  Moreover, the scaling of
Newton's coupling \eqref{cl_scaling_g} allows us to identify a scale
in the following way. As $g$ (or $\overline g$) scales classically,
the coefficient of proportionality, say $C$, is nothing else than the
Newton constant, because
\begin{equation}
 G_N = g k^{-2} = C k^2 k^{-2} = C \, .
\end{equation}
Thus, scales are measured in units of the Planck mass, $M_{Pl}^2 =
1/C$. The physical trajectory is then fixed by measuring the relevant
couplings, that is the (background) Newton constant and cosmological
constant, at a given scale. In \autoref{fig:traj} one can see both,
the classical scaling in the \IR{} as well as the vanishing of the
$\beta$-functions in the vicinity of the \UV{} fixed point for large
$k$.  The classical scaling regime extends roughly up to one order of
magnitude below the Planck scale. This implies the absence of quantum
gravity effects for energies $E \ll M_{\mathrm{Pl}}$, as it is
expected in a theory without a large volume compactification of
extra-dimensions.  Note also that the difference between the two
couplings $g$ and $\bar g$ is hardly visible, which justifies to some
extent the background approximation for the Newton constant. \\[-.5ex]

\section{Summary and outlook}

We have presented a quantum gravity calculation that shows a classical
regime on large distances, and asymptotically safe physics in the
non-perturbative \UV{} limit. The behaviour at large distances
relates to an attractive \IR{} fixed point with classical scaling
behaviour. This implies that the dimensionful Newton constant $G_N$ and
the cosmological constant $\Lambda$ do not depend on the energy scale
(inverse length scale) for large distances. Hence, for the first time,
this includes the domain of classical gravity, that has been tested
experimentally, in the renormalization group approach to quantum
gravity. The classical gravity regime in the vicinity of the \IR{}
fixed point is connected to a non-perturbative \UV{} fixed
point, which ensures the finiteness of scattering amplitudes at arbitrary
high energies. The small dependence of the results on the
regulator indicates stability of the present truncation. Technically,
this work introduces a novel approximation scheme in \RG{}-gravity
calculations. This scheme has two essential features: First, we work in a vertex
expansion with fully momentum-dependent wave-function renormalizations
for the graviton and for the ghost field. Second, the higher order correlation
functions are parametrized by additional couplings for their
momentum-dependent and momentum-independent parts. The latter become
important in the \IR{} and their properties are determined by a
self-consistency scaling analysis. In particular, we show that the
momentum-independent part of the two point function cannot be
identified with the cosmological constant at large distances.

In summary, this work provides further evidence for the asymptotic
safety scenario in quantum gravity. In addition it substantiates the
physics at the \IR{} fixed point found in
\cite{Donkin:2012ud,Christiansen:2012rx}. By now, the approximation is
quantitative enough to produce classical scaling for the couplings
$G_N$ and $\Lambda$ for large length scales, in accordance with
experimental observations. The present approximation is readily
extended to include higher correlation functions which will be
reported on in future work \cite{Vertices}.

\section*{Acknowledgments} We thank Tobias Henz, Daniel F.~Litim and
Christof Wetterich for discussions and collaboration on related
subjects. We thank Alessandro Codello for providing us with an updated version
of \cite{Codello2013} prior to substitution on arXiv. JMP thanks the Yukawa
Institute for Theoretical Physics,
Kyoto University, where this work was completed during the
YITP-T-13-05 on 'New Frontiers in QCD'. This work is supported by
Helmholtz Alliance HA216/EMMI and by ERC-AdG-290623.

\appendix
\allowdisplaybreaks
\section{Flow equations} \label{app_flow_eq}

The modified threshold functions introduced in \autoref{sec_gdot} are given by:
\begin{equation}
 \Phi_n^p[\eta](\omega) = \frac{1}{\Gamma(n)} \int\limits_0^\infty \text{d}x x^n
\frac{\dot r(x) - \eta(x) r(x)}{(x(1+r(x))+\omega)^p} \, .
\end{equation}
With these threshold functions, the geometric flow equations in
\cite{Donkin:2012ud}, improved by momentum-dependent
anomalous dimensions, are given by
\begin{align}
 \notag \dot{\overline g} = 2\overline{g} - \frac{\overline{g}^2}{2\pi} \bigg[
\frac{2}{3} \Phi_1^1[\eta_h](\mu) + \frac{10}{3} \Phi_2^2[\eta_h](\mu)& \\
+\frac{5}{12} \Phi_1^1[\eta_c](0) + \frac{5}{4} \Phi_2^2[\eta_c](0) \bigg]& \, ,
\\
 \notag \dot g = 2g - \frac{g^2}{2\pi} \bigg[ \frac{2}{3} \Phi_1^1[\eta_h](\mu)
+ \frac{10}{3} \Phi_2^2[\eta_h](\mu)& \\ \notag +\frac{5}{24}
\Phi_1^1[\eta_c](0) + \frac{5}{8} \Phi_2^2[\eta_c](0)& \\ +\dot\mu \left(
\frac{2}{3} \Phi_1^2[0](\mu) + \frac{20}{3} \Phi_2^3[0](\mu) \right) \bigg]& \,
.
\end{align}
The flows of the two-point functions are given by
\begin{align}
 &\frac{\partial_t \Gamma^{(2h)}(p^2)}{Z_h(p^2)} = g \int\limits_0^\infty
\text{d}q \int\limits_{-1}^1 \text{d}x \sqrt{1-x^2} \frac{q^3}{3\pi^2} \times 
\notag \\
&\qquad\qquad\quad \Bigg[ -\frac{f_c(p,q,x) (\dot r_1 - \eta_c(q^2)
r_1)}{(1+r_1)^2(1+r_2)} \\ 
&\qquad\qquad\quad -\frac{6q^2
(3p^2+6q^2-8\lambda^{(4)}) (\dot
r_1-\eta_h(q^2) r_1)}{(q^2(1+r_1)+\mu)^2} \notag \\ 
&+ \frac{f_h(p,q,x,\lambda^{(3)})
(\dot r_1 -
\eta_h(q^2) r_1)}{(q^2(1+r_1)+\mu)^2((p^2+2pqx+q^2)(1+r_2)+\mu)} \Bigg] \, , 
\notag \\
 &\frac{\partial_t \Gamma^{(\overline{c}c)}(p^2)}{Z_c(p^2)} = -g
\int\limits_0^\infty \text{d}q \int\limits_{-1}^1 \text{d}x \sqrt{1-x^2}
\frac{4q^3}{3\pi^2} \times \notag \\
 &\frac{p^2(3+6x^2) + pqx(-2+20x^2) +
q^2(5-12x^2+16x^4)}{p^2+2pqx+q^2} \times \\ 
 &\Bigg[ \frac{\dot r_1 -
\eta_c(q^2)r_1}{(1+r_1)^2((p^2+2pqx+q^2)(1+r_2)+\mu)} \notag \\ 
&\qquad\qquad\qquad\qquad\qquad +\frac{q^2(\dot r_1 -
\eta_h(q^2)r_1)}{(q^2(1+r_1)+\mu)^2(1+r_2)} \Bigg] \, .\notag
\end{align}
Here, we have introduced the short cuts $r_1 = r(q^2)$ and $r_2 = 
r(p^2+2pqx+q^2)$.
The functions $f_c(p,q,x)$ and $f_h(p,q,x,\lambda^{(3)})$ read
\begin{equation}
 f_c(p,q,x) = \frac{16\left(p^2(2+x^2)+6pqx+3q^2\right)}{p^2+2pqx+q^2}
\end{equation}
and
\begin{align}
 f_h(p&,q,x,\lambda^{(3)}) = \frac{4q^2}{5(p^2+2pqx+q^2)} \times \notag \\ 
\Big(&15p^6(1+2x^2)+10p^5qx(7+8x^2) \notag \\ 
&+p^4q^2(21+208x^2+56x^4)+140p^3q^3x(1+2x^2) \notag \\
 &+4p^2q^4(7+76x^2+22x^4) \notag \\
 &+8pq^5x(17+11x^2+2x^4)  \notag \\
 &+4q^6(7+6x^2+2x^4) - 4\lambda^{(3)}\big[ 15p^4(1+4x^2) \notag  \\ 
&+30p^3qx(3+4x^2)+p^2q^2(-9+248x^2+16x^4) \notag \\
&+20pq^3x(5+4x^2)+20q^4(1+2x^2)\big] \notag \\
 &+8(\lambda^{(3)})^2\big[15p^2(1+x^2)+20pqx(2+x^2) \notag \\
 &+2q^2(9+2x^2+4x^4)\big] \Big) \, .
\end{align}
Finally, the flow equation for $\lambda$ from the one-point function is given by
\begin{align}
 \dot\lambda =& -2\lambda + g\Bigg[\int\limits_0^\infty \text{d}q
\frac{3q^7(1+2\lambda^{(3)})}{4\pi} \frac{\dot r(q^2) - \eta_h(q^2)
r(q^2)}{(q^2(1+r(q^2))+\mu)^2} \notag \\
 &+ \int\limits_0^\infty \text{d}q
\frac{q^3(3-\lambda^{(3)})}{3\pi} \frac{\dot r(q^2) - \eta_c(q^2)
r(q^2)}{(1+r(q^2))^2} \Bigg] \, .
\end{align}

\section{Projection procedure}\label{app_anomproj}

Here, we want to argue why the projection procedure to obtain the
flow equation for the mass and the integral equation for the graviton anomalous
dimension is the physical one. First the anomalous
dimension must be finite everywhere, in particular at the pole. Otherwise, the
wave-function renormalization could be written as
\begin{equation}
 Z_h(p^2) = (p^2+\mu)^\omega \tilde Z_h(p^2)
\end{equation}
with a nonzero parameter $\omega$ and $Z_h(-\mu)$ finite and nonzero. However,
the wave-function renormalization should only determine the residue at the
propagator pole. Thus, we assume that the anomalous dimension is finite.

Next, we consider the flow equation of the two-point function,
\begin{equation}
 -\eta_h(p^2)(p^2+\mu) + \dot\mu +2\mu = \frac{\partial_t
\Gamma^{(2h)}(p^2)}{Z_h(p^2)} \, .
\end{equation}
We can evaluate this equation at an arbitrarily chosen fixed momentum, say 
$\ell$, to
obtain the $\beta$-function of the mass:
\begin{equation}
 \dot\mu = -2\mu + \frac{\partial_t\Gamma^{(2h)}(\ell^2)}{Z_h(\ell^2)} +
\eta_h(\ell^2)(\ell^2+\mu) \, .
\end{equation}
Substracting this from the original equation leaves an integral equation for
the anomalous dimension,
\begin{equation}\label{eq_etawithproj}
 \eta_h(p^2) =
-\frac{\frac{\partial_t\Gamma^{(2h)}(p^2)}{Z_h(p^2)}-\frac{\partial_t\Gamma^{
(2h)}(\ell^2)}{Z_h(\ell^2)}}{p^2+\mu} + \eta_h(\ell^2)
\frac{\ell^2+\mu}{p^2+\mu} \, .
\end{equation}
One easily sees that the right hand side of this equation diverges at
$p^2=-\mu$, if $\eta(\ell^2)$ is not chosen appropriately. As we already know
that the anomalous dimension must be finite everywhere, we conclude that
\begin{equation}
 \eta_h(\ell^2) =
\frac{\frac{\partial_t\Gamma^{(2h)}(-\mu)}{Z_h(-\mu)}-\frac{\partial_t\Gamma^{
(2h)}(\ell^2)}{Z_h(\ell^2)}}{\ell^2+\mu} \, .
\end{equation}
This can be reinserted into \eqref{eq_etawithproj}, leading to
\begin{equation}
 \eta_h(p^2) =
-\frac{\frac{\partial_t\Gamma^{(2h)}(p^2)}{Z_h(p^2)}-\frac{\partial_t\Gamma^{
(2h)}(-\mu)}{Z_h(-\mu)}}{p^2+\mu} \, ,
\end{equation}
which is the equation originally proposed in the main text. We can use the
expression for $\eta_h(\ell^2)$ also for the flow equation of the mass,
resulting in
\begin{equation}
 \dot\mu = -2\mu + \frac{\partial_t\Gamma^{(2h)}(-\mu)}{Z_h(-\mu)} \, ,
\end{equation}
again reproducing the result from the main text. Thereby, we have shown,
under the reasonable condition of a finite anomalous dimension, that our flow
equations are unique.

\section{Infrared scaling analysis} \label{appendix_analysis}

Here, we discuss the \IR{} divergence analysis in more
detail. Before we can proof a recursion relation for the parameters $\alpha_n$,
we discuss some general properties of such a setting.

\textbf{Prerequisites:}
The flow of a general $n$-point vertex function includes generic loop integrals
with dimensionless external momenta $p_i$ and loop momentum $q$ of the form 
\begin{equation}
\begin{aligned}
\int &\frac{\mathrm{d}^4 q}{(2 \pi)^4} \frac{f_{n}(q,p_i)}{(q^2(1+r(q^2)) +
\mu)^2} \times \\
&\prod\limits_{i=1}^{n-2}((p_i+q)^2(1+r((q+p_i)^2))+\mu)^{-1}\, ,
\end{aligned}
\end{equation}
with a function $f_n(q,p_i)$ resulting from contractions of the tensor
structure $\tilde{\mathcal{T}}$ defined in \eqref{eq_lambda_n}. Obviously, the
divergences are strongest at vanishing
external momenta $p_i=0$. In this case, all internal propagators carry the loop
momentum $q$ and we are left with 
\begin{equation}\label{eq_div_integral}
\int \frac{\mathrm{d}^4 q}{(2 \pi)^4} \frac{f_n(q,p_i=0)}{(q^2(1+r(q^2)) +
\mu)^n
} \, .
\end{equation}
Moreover, in the limit $\mu \rightarrow -1$, these divergences emerge from
small momentum modes near $q=0$. Consistent regulators need to fulfill
\begin{equation}
\lim_{x\rightarrow 0} x(1+ r(x)) = 1 + \zeta x + \mathcal{O}(x^2)   
\end{equation}
with $\zeta > 0$, because otherwise, the denominator in \eqref{eq_div_integral}
exhibits a zero for $\mu > -1$.  
The highest pole order is contained in the momentum-independent part
$f_n^0 \colonequals f_n(q=0,p_i=0)$. 
Neglecting the angular integration, and
by the above reasoning, the most divergent
part of the integral takes the form
\begin{equation}
\int_0^{\delta>0} \mathrm{d}q \frac{q^3 f_n^0}{(q^2 + \epsilon)^n
} \, ,
\end{equation}
where we introduced $\epsilon = 1+\mu$ for convenience.
These expressions can be integrated which leads to a divergence structure of
the form
\begin{equation}
\int_0^{\delta>0} \mathrm{d}q \frac{q^3}{(q^2 + \epsilon)^n
} \sim \begin{cases}
        \mathrm{finite} \hspace{5pt} \mathrm{if} \hspace{5pt} n<2
	\\ \log \epsilon \hspace{8pt} \mathrm{if} \hspace{5pt} n=2
	\\ \epsilon^{2-n} \hspace{7pt} \mathrm{if} \hspace{5pt} n>2
       \end{cases} \, .
\label{integrals}\end{equation}
In a simple Einstein Hilbert truncation one identifies the constant,
momentum-independent parts of the $n$-point vertex functions $\lambda^{(n)}$ with
the mass
term (or with the cosmological
constant), i.\,e.\ $\lambda^{(n)} = -\mu/2$ for all $n$. However, this
approximation
incorporates a scaling inconsistency, as we will see by the divergence
analysis
below.

\begin{figure}
\includegraphics[width=\columnwidth]{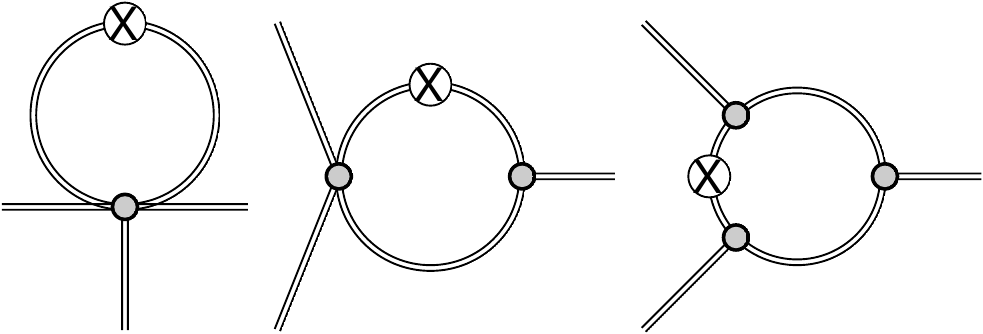}
\caption{Diagrams contributing to the divergence analysis of the three-point
function of the graviton.}
 \label{fig:threepointdiags}
\end{figure}

The following analysis is based on a matching of terms in the
limit
$\epsilon \rightarrow 0$ on the RHS and the LHS of the flow equations for
$n$-point vertex functions. As a generalization of the Einstein-Hilbert
construction we allow for a power law behaviour in $\epsilon$ in the limit under
consideration. In such an expansion, logarithmic contributions are sub-leading,
do
not change the power law and are therefore discarded. Moreover, we keep only
the leading order terms, i.e.\ we
assume a power law 
\begin{equation}
\lambda^{(n)} \stackrel{\epsilon \rightarrow 0}{\sim} \epsilon^{\alpha_n}\, ,
\,\, n\geq 3 \, ,
\label{ansatz_divstruc0}\end{equation}
and suppress terms of the form $\epsilon^{\tilde{\alpha}_n}$ with
$\tilde{\alpha}_n > \alpha_n$, since we are interested in the limit $\epsilon
\rightarrow 0$.
Moreover, we observe that
the $\epsilon$-dependence of the function $f_n^0$ is completely stored in the
parameters $\lambda^{(n)}$. Accordingly, these are the only terms that we have
to take into account in an $\epsilon$-scaling analysis. 
The generic form of a $\beta$-function for $\lambda^{(n)}$ is of the form
\begin{equation}
\dot{\lambda}^{(n)} = - 2 \lambda^{(n)} + g \,\, (\mathrm{loop-terms}) \, , 
\end{equation}
i.e.\ a canonical term and loop contributions which are always proportional to
the gravitational coupling $g$. First, we show that the canonical term does not
dominate the $\beta$-functions for $\lambda^{(n)}$ with $n=2,3,4$ in the limit
$\epsilon \rightarrow 0$. 
In order to do so, we assume that the canonical term in the beta functions for
$\lambda^{(3)}$ dominates in the limit $\epsilon \rightarrow 0$, i.e.\
$\dot{\lambda}^{(3)} \stackrel{\epsilon \rightarrow 0}{\sim} - 2
\lambda^{(3)}$, which implies $\lambda^{(3)} \stackrel{\epsilon \rightarrow
0}{\sim} 1/k^2$. On the other hand, dominance of the canonical term means
$\dot{\lambda}^{(3)} \stackrel{\epsilon \rightarrow 0}{\sim}
\epsilon^{\alpha_3}$. From \autoref{fig:threepointdiags} and
\eqref{integrals} we can can see that there is a diagram producing a term
$\stackrel{\epsilon \rightarrow 0}{\sim} \epsilon^{3 \alpha_3 -2}$. Dominance
of the canonical term then implies $\alpha_3 >1$. In this case $\lambda^{(3)}
\stackrel{\epsilon \rightarrow 0}{\rightarrow} 0$. This contradicts
$\lambda^{(3)} \stackrel{\epsilon \rightarrow 0}{\sim} 1/k^2$ as long as there
is no \UV{} \, fixed point at $\epsilon = 0$. 
The same argument goes through for $\lambda^{(4)}$ by using the term
$\stackrel{\epsilon \rightarrow 0}{\sim} \epsilon^{2 \alpha_4 -1}$ on the RHS
of the respective flow equation. Thus, we know that the canonical term is
sub-leading or of equal order as the loop terms.
With a case-by-case analysis of $\left( \lambda_3 \lessgtr 1, \lambda_4\lessgtr
1 \right)$, one can show that $\alpha_4 < 1$.
Then, using \eqref{RHS_lambda4} below it can be deduced that the canonical term
in the flow equation for $\lambda^{4}$ is indeed subleading and
$\dot{\lambda}^{4} \stackrel{\epsilon \rightarrow 0}{\sim} g$. Together with
\eqref{LHS_lambda4} this in turn
implies that $\dot{\epsilon} \stackrel{\epsilon \rightarrow 0}{\sim} g$ and
therefore, the canonical term in
$\dot{\epsilon}$ is irrelevant as well. Accordingly either
\begin{equation}
\alpha_4 <0 \hspace{7pt} \mathrm{or} \hspace{7pt} \alpha_3 < 1/2 \, , 
\label{eq_gen_cond}\end{equation}
and additionally, the canonical term in $\dot{\lambda}^{n}$
for all $n$ is sub-leading too, as both sides of the respective flow
equation must be proportional to $g$. 

\textbf{Lemma 1:} \textit{Assuming the results in Prerequisites, in particular
a power law
\begin{equation}
\lambda^{(n)} \stackrel{\epsilon \rightarrow 0}{\sim} \epsilon^{\alpha_n}\,,   
\label{ansatz_divstruc}\end{equation}
with $n\geq 3$, the hierarchy of
flow equations implies
\begin{equation}
\alpha_4 \leq 2 \alpha_3 -1 \, . 
\label{final_inequality}\end{equation}}

\begin{figure}
\includegraphics[width=\columnwidth]{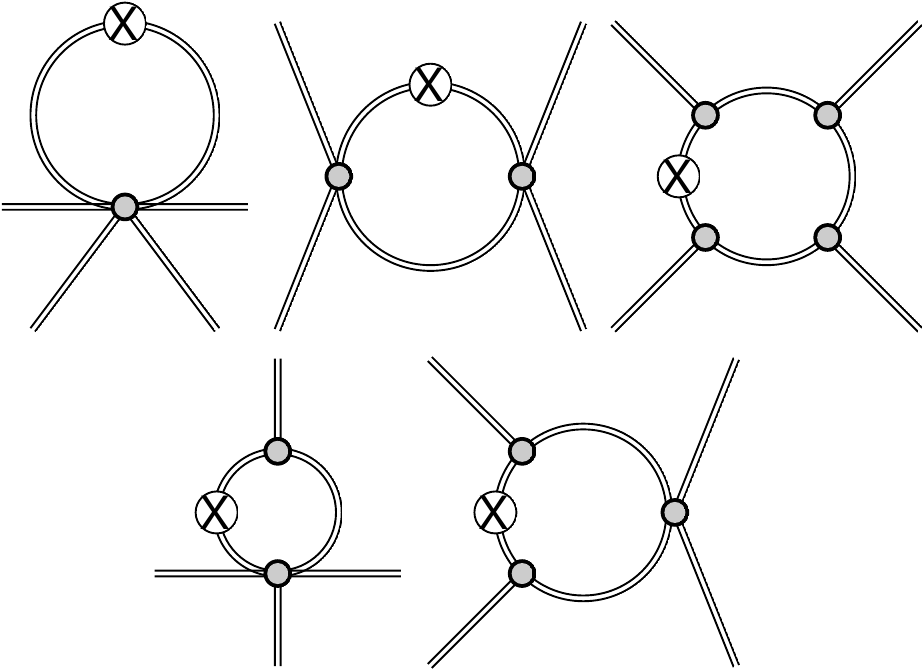}
\caption{Diagrams contributing to the divergence analysis of the four-point
function of the graviton.}
 \label{fig:fourpointdiags}
\end{figure}

\textit{Proof}: The flow of the two point function (evaluated at vanishing
external momentum) leads to the relation
\begin{equation}\label{mu_divstruc}
\begin{aligned}
\dot{\mu} = \dot{\epsilon} \stackrel{\epsilon \rightarrow 0}{\sim} &g \,
\max\left\{\left|\lambda^{(4)}\right|,\frac{(\lambda^{(3)})^2}{\epsilon}\right\}
 \\ \sim &g  \, \max \left\{\epsilon^{\alpha_4},\epsilon^{2 \alpha_3 -1}\right\}
\, ,
\end{aligned}
\end{equation}
where the terms in the curly brackets $\{\cdot,\cdot,...\}$ indicate
the leading contributions arising from
the distinct diagrams.
The diagrams generating the running of the three point function (see
\autoref{fig:threepointdiags}) lead to
\begin{equation}
\dot\lambda^{(3)} \stackrel{\epsilon \rightarrow 0}{\sim} 
g  \, \max \left\{\epsilon^{\alpha_5}
,\epsilon^{\alpha_3 + \alpha_4
-1},\epsilon^{3\alpha_3 - 2}\right\} \,.\label{RHS_lambda3}\end{equation}
The next order in the hierarchy, the flow equation for $\Gamma^{(4h)}$, has the
diagrammatic representation \autoref{fig:fourpointdiags}.
This means that the diagrams scale in the limit $\epsilon \rightarrow 0$ as
\begin{equation}\label{RHS_lambda4}
\begin{aligned}
\dot\lambda^{(4)} \stackrel{\epsilon \rightarrow 0}{\sim} g  \, \max
\big\{ &\epsilon^{4 \alpha_3-3},\epsilon^{2
\alpha_3+\alpha_4-2}, \\ &\epsilon^{2 \alpha_4
-1},\epsilon^{\alpha_3+\alpha_5-1},\epsilon^{\alpha_6} \big\}.
\end{aligned}
\end{equation}

On the other hand, we can calculate $\dot\lambda^{(3)}$ and $\dot\lambda^{(4)}$ 
from \eqref{ansatz_divstruc}, i.e.\ the LHS of the flow equation.
Using \eqref{mu_divstruc}, this yields 
\begin{equation}
\dot\lambda^{(3)} \stackrel{\epsilon \rightarrow 0}{\sim} 
\epsilon^{\alpha_3-1}
\dot{\epsilon}
\stackrel{\epsilon \rightarrow 0}{\sim}
g  \, \max \left\{\epsilon^{\alpha_3+\alpha_4-1} , \epsilon^{3 \alpha_3
-2}\right\} \, ,
\label{LHS_lambda3}\end{equation}
and
\begin{equation}
\dot\lambda^{(4)} \stackrel{\epsilon \rightarrow 0}{\sim} 
\epsilon^{\alpha_4-1}
\dot{\epsilon} \stackrel{\epsilon \rightarrow 0}{\sim}
g  \, \max \left\{\epsilon^{2\alpha_4-1}, \epsilon^{\alpha_4+2
\alpha_3-2}\right\} \, .
\label{LHS_lambda4}\end{equation}
Consistency requires matching of the leading terms on both
sides of the flow equations.   

If the leading term on the LHS is known, the
matching condition induces inequalities on the RHS and we can obtain relations
for $\alpha_n$. 

On the left hand sides of the flow equations for
the three- and the four-point functions we have in each case two terms,
\eqref{LHS_lambda3} and \eqref{LHS_lambda4}. Let us now study the different
cases
for the leading terms.
Considering the three-point function, the different cases can be written
as
\begin{align}
\alpha_3 + \alpha_4 -1 & \lesseqqgtr 3 \alpha_3 -2  \, .
\label{lambda_3_LHS_rel}\end{align}
Analogously, from the LHS of the four-point function we obtain one of the
``dual'' relations
\begin{align}
2 \alpha_4 -1 & \lesseqqgtr \alpha_4 + 2 \alpha_3 -2 \, .
\label{lambda_4_LHS_rel}\end{align}
Obviously, the above relations \eqref{lambda_3_LHS_rel} and
\eqref{lambda_4_LHS_rel} are equivalent, i.e.\ if one of the relations
$>,=,<$ is true, the corresponding relation holds for the other
expression. 
Consequently, if
we know the leading
term on the LHS of the flow equation for $\Gamma^{(3h)}$, we know the leading
term in the corresponding equation for $\Gamma^{(4h)}$ and vice versa. We
proceed with a case-by-case analysis.

\textit{(i) Assume 
\begin{equation}
\alpha_3 + \alpha_4 -1 \geq 3 \alpha_3 -2  \, .
\label{case_1}\end{equation}}
From the ``dual'' relation for the four point function we know that $\alpha_4 +
2
\alpha_3 -2$ is the dominant term on the LHS of $\dot{\lambda}^{(4)}$ and
therefore also on the RHS, i.e.\ in \eqref{RHS_lambda4}. Hence, the inequality
\begin{equation}
\alpha_4 + 2 \alpha_3 -2 \leq 4 \alpha_3 -3 
\end{equation}
necessarily holds. This inequality in turn implies $\alpha_4 \leq 2 \alpha_3
-1$, while \eqref{case_1} is equivalent to $\alpha_4 \geq 2 \alpha_3-1$.
We conclude that
\begin{equation}
\alpha_4 = 2 \alpha_3-1 \, ,
\label{x}\end{equation}
which is equivalent to the assumption $3 \alpha_3 -2 = \alpha_3 + \alpha_4 -1$
while $3 \alpha_3 -2 < \alpha_3 + \alpha_4 -1$ produces a contradiction.

With \eqref{x} we have checked two of the three cases, i.e.\ under the $\geq$
assumption only the $=$ sign is a consistent solution.
The term involving $\alpha_5$ is subject to the condition
\begin{equation}
\alpha_5 \geq \alpha_3 + \alpha_4 -1 \,.
\end{equation}
Let us study the other case:

\textit{(ii) Assume
\begin{equation}
\alpha_3 + \alpha_4 -1 < 3 \alpha_3 -2  \, .
\label{case_2}\end{equation}}
Comparing with \eqref{RHS_lambda3}, we find that the
equation for $\dot{\Gamma}^{(3h)}$ is trivially consistent with this assumption.
Moreover, the assumption that $2\alpha_4 -1$ is the leading term (which is
equivalent to assumption \eqref{case_2}) is consistent with the first three
diagrams in \eqref{RHS_lambda4}. Again, the terms involving $\alpha_5$ and
$\alpha_6$ are appropriately constrained, see remark 2 below, and we can
constitute that
\eqref{case_2} is indeed a consistent assumption. Including both cases, this
leads to the relation
\begin{equation}
\alpha_4 \leq 2 \alpha_3 -1 \, , 
\end{equation}
and proves the lemma. $\Box$ \\

\textit{Remark 1:} This means that from $\dot{\Gamma}^{(3h)}$ and
$\dot{\Gamma}^{(4h)}$ we obtain an inequality that constrains the relation
between $\alpha_3$ and $\alpha_4$, which are at this stage free parameters. \\

\textit{Remark 2:} Lemma 1 with \eqref{eq_gen_cond} implies 
\begin{equation}
\alpha_4 < 0 
\label{eq_alpha4_constraint}\end{equation}
always, thus $\alpha_3$
cannot be constrained solely by the analysis above. \\

\textit{Remark 3:} Moreover, we cannot fix $\alpha_n$ for $n>4$ with the
equations for the three- and the four-point function. However, since equations
\eqref{LHS_lambda3} and \eqref{LHS_lambda4} are independent of $\alpha_4$ and
$\alpha_5$, these terms cannot be the leading contributions in the
limit under consideration, i.e.\ they cannot generate the power law. This
implies
\begin{equation}
\alpha_5 \geq \alpha_4 + \alpha_3 -1 \,. 
\label{bound_lambda5}\end{equation}
Applying the same logic to the four-point function, we arrive at
\begin{equation}
\alpha_6 \geq 2 \alpha_4 -1 \,. 
\label{bound_lambda6}\end{equation}

In order to further constrain the parameters $\alpha_n$ with $n>4$, we proceed
by analyzing the
running of $\Gamma^{(nh)}$. 
\begin{table*}
\begin{center}
\begin{tabular}{|c||c|c|c|c|c|} \hline
 $1/\chi$ & $\alpha_3=-0.1$ & $\alpha_3=-0.2$ & $\alpha_3=-0.3$ &
$\alpha_3=-0.4$ & $\alpha_3=-0.5$ \\
\hline \hline
 10 & $\mu_*=-0.645$ & $\mu_*=-0.651$ & $\mu_*=X$ & $\mu_*=X$ & $\mu_*=X$ \\
  & $g_*=0.614$ & $g_*=0.604$ & $g_*=X$ & $g_*=X$ & $g_*=X$ \\ \hline
 15 & $\mu_*=-0.630$ & $\mu_*=-0.633$ & $\mu_*=-0.637$ & $\mu_*=-0.641$ &
$\mu_*=X$ \\
  & $g_*=0.637$ & $g_*=0.633$ & $g_*=0.628$ & $g_*=0.620$ & $g_*=X$ \\ \hline
 20 & $\mu_*=-0.624$ & $\mu_*=-0.626$ & $\mu_*=-0.628$ & $\mu_*=-0.630$ &
$\mu_*=-0.634$ \\
  & $g_*=0.647$ & $g_*=0.644$ & $g_*=0.640$ & $g_*=0.637$ & $g_*=0.632$ \\
\hline
 25 & $\mu_*=-0.621$ & $\mu_*=-0.622$ & $\mu_*=-0.623$ & $\mu_*=-0.625$ &
$\mu_*=-0.627$ \\
  & $g_*=0.653$ & $g_*=0.651$ & $g_*=0.648$ & $g_*=0.645$ & $g_*=0.642$ \\
\hline
 30 & $\mu_*=-0.618$ & $\mu_*=-0.619$ & $\mu_*=-0.621$ & $\mu_*=-0.620$ &
$\mu_*=-0.623$ \\
  & $g_*=0.658$ & $g_*=0.655$ & $g_*=0.653$ & $g_*=0.651$ & $g_*=0.648$ \\
\hline
 35 & $\mu_*=-0.617$ & $\mu_*=-0.618$ & $\mu_*=-0.619$ & $\mu_*=-0.620$ &
$\mu_*=-0.621$ \\
  & $g_*=0.659$ & $g_*=0.658$ & $g_*=0.657$ & $g_*=0.655$ & $g_*=0.652$ \\
\hline
\end{tabular}
\caption{\UV{} fixed point values for different parameters $\chi$ and $\alpha_3$
with exponential regulator and $a=4$. $X$ indicates that no fixed point is found
for these parameter
values.}
\label{tab:UVFPparamscan}
\end{center}
\end{table*}

\textbf{Lemma 2:} \textit{Under the same conditions as in Lemma
1, we obtain the recursion relation for $n\geq5$:
\begin{equation}
\alpha_n = \alpha_{n-2} + \alpha_4 -1 \,.
\end{equation}
}

\textit{Proof}: In general, assumption \eqref{ansatz_divstruc} together with
\eqref{mu_divstruc} leads to
\begin{equation}
\begin{aligned}
\dot{\lambda}^{(n)}  \stackrel{\epsilon \rightarrow 0}{\sim}& g 
\,\epsilon^{\alpha_n-1} \dot{\epsilon}  \\
 \stackrel{\epsilon \rightarrow 0}{\sim}& g
\max \left\{ \epsilon^{\alpha_n + \alpha_4 -1}, \epsilon^{\alpha_n +2 \alpha_3
-2}
\right \} \, .
\end{aligned}\
\end{equation}
The canonical term can be dropped since $\alpha_4 < 0$.
Lemma 1 ensures that the leading term is always given by $\epsilon^{\alpha_n +
\alpha_4 -1}$. Accordingly, all terms generated by the diagrams on the RHS of
the flow equation must be smaller or equal to this term. For every $n$, the flow
equation
for $\Gamma^{(nh)}$ contains a diagram with $2$ four-point
vertices, one $(n-2)$-vertex and $4$ internal propagators. Hence, we can
conclude
\begin{equation}
\alpha_n+ \alpha_4 -1 \leq 2 \alpha_4 +\alpha_{n-2} -2 \,.
\label{inequality_1}\end{equation}
Moreover, we can generalize the results \eqref{bound_lambda5} and
\eqref{bound_lambda6}, by considering the diagram with only one vertex, more
precisely, an $(n+2)$-vertex. Again, consistency requires
\begin{equation}
\alpha_{n+2} \geq \alpha_n + \alpha_4 -1 \, 
\end{equation}
or equivalently $\alpha_n \geq \alpha_{n-2} + \alpha_4 -1$.
Combining this result with \eqref{inequality_1} proves lemma 2. $\Box$ \\

\textit{Remark 4:}
Lemma 2 yields a recursion relation connecting all $\alpha_n$ with
$\alpha_4$ and $\alpha_3$.
We are therefore left with two free parameters, with the
constraint \eqref{eq_alpha4_constraint}.
Moreover, we can give explicit, non-recursive, expressions for $\alpha_n$,
depending on whether $n$ is odd or even. The difference $\Delta \alpha$ between
$\alpha_n$ and $\alpha_{n-2}$ is obviously
\begin{equation}
\Delta \alpha = \alpha_4 -1 \, , 
\end{equation}
and therefore independent of $n$. For $n$ even we can express any $\alpha_n$ as
\begin{align}
\notag \alpha_n =&  \alpha_4 + \left( \frac{n-4}{2} \right)  \Delta \alpha 
\\ =&
\left( \frac{n}{2} -1 \right) \alpha_4 - \left( \frac{n}{2}-2 \right) \, , \,
n \, \mathrm{even} \, ,n \geq 6  \, ,
\end{align}
which can be rewritten as
\begin{equation}
\alpha_{2n} = (n-1)\alpha_4-(n-2) \, , \, \, n \geq 3 \,. 
\end{equation}
For $n$ odd we can do the same thing, starting with $\alpha_3$ and adding
multiples of $\Delta \alpha$ in order to arrive at
\begin{equation}
\alpha_{2n+1} = \alpha_3 + (n-1)\alpha_4 -(n-1)\, , \, \, n \geq 2 \,. 
\end{equation}

\section{Estimating \texorpdfstring{$\alpha_3$}{alpha3}}
\label{appendix_estimate}

Following the argument in \autoref{sec_lambdas}, there is a transition regime
between
Einstein-Hilbert-type of solutions with $\lambda^{(2)} = \lambda^{(3)}$ and an
\IR{} regime where they differ as the trajectories approach $\epsilon =0$.
The simplest form of such a transition is a sharp switch between the two
solutions at a scale $k_0$ when the trajectories bend over to the separatrix
and are attracted towards the \IR{} fixed point. 
Such a sharp cross-over between these two regions is certainly different from 
the true behaviour, and it also differs from our ansatz \eqref{c_n_param}.
However, this type of transition shares the essential features with the true 
solution and can thus provide a reasonable estimate.
At the transition scale $k_0$ we have the connection conditions
\begin{equation}
\lambda^{(3)} (k_0) = \lambda^{(2)} (k_0) = - \frac{1}{2} \mu(k_0)  
\label{connection_cond_1}\end{equation}
and
\begin{equation}
\dot{\lambda}^{(3)} (k_0) = \dot{\lambda}^{(2)} (k_0) = - \frac{1}{2}
\dot{\epsilon}(k_0) \, . 
\label{connection_cond_2}\end{equation}
Using the power law \eqref{ansatz_divstruc0} we have the simple
relation for the logarithmic derivative
\begin{equation}
\frac{\left(\lambda^{(3)}\right)'}{\lambda^{(3)}} = \frac{\alpha_3}{\epsilon}
\, ,
\label{log_der}\end{equation}
where $'$ denotes the derivative with respect to $\epsilon$. The scale
derivative can be expressed as
\begin{equation}
\dot{\lambda}^{(3)} = \left(\lambda^{(3)}\right)' \dot{\epsilon} \, .
\end{equation}
Evaluating the above equation at $k=k_0$ and combining it with
\eqref{connection_cond_2} yields
\begin{equation}
\left. \left(\lambda^{(3)}\right)' \right|_{k=k_0} = - \frac{1}{2} \, . 
\end{equation}
This in turn can be used together with \eqref{connection_cond_2} when
evaluating \eqref{log_der} at $k=k_0$. Keeping in mind that $\epsilon = 1 +
\mu$,
this results in
\begin{equation}
\alpha_3 = \frac{1 + \mu(k_0)}{\mu(k_0)} \,. 
\end{equation}
Together with $\alpha_4 = 2\alpha_3-1$ this fixes all $\alpha_n$.
From the phase diagram, one infers that the onset of this transition is near
$\mu \approx -0.9$ independent of the chosen parameters. This gives
\begin{equation}
 \alpha_3 \approx -1/9 \, .
\end{equation}
Not sticking to the equality does not alter the results qualitatively.

\section{Functional form of
\texorpdfstring{$\lambda^{(n)}$}{lambdan}}\label{app_funcform}

Here we construct explicit expressions for the functions $\lambda^{(n)}$. In
addition to the singularity structure, there are further constraints to be
fulfilled by such an ansatz. In the following we show that
\begin{equation}\label{c_n_param_app}
\begin{aligned}
\lambda^{(n)} &= - \frac{\mu}{2} \left[ 1 + \mathrm{sgn}(\mu) \, \chi \left|
\frac{\mu}{1+\mu}\right|^{-\alpha_n}\right] \\
&=- \frac{\mu}{2} (1 + \delta \lambda^{(n)}) 
\end{aligned}
\end{equation}
is consistent with all constraints. In the above formula $\chi$ is an arbitrary
parameter. From perturbation theory we know that in the Gaussian limit
$\mu \rightarrow 0$, we need to recover an Einstein-Hilbert solution.
Indeed, \eqref{c_n_param_app} entails $\lambda^{(n)} = -\mu/2$ for all $n$ in
the vicinity of $\mu=0$ as long as $\alpha_n < 0$. In addition to that,
it is clear that the correction
cannot contain further powers of $g$, since this would interfere with the
singularity structure. Furthermore, the correction should be inherently
dimensionless. With these conditions, a quantity proportional to powers of the
ratio $\mu / (1+\mu)$ is everything we have at hand. In the end, we are left 
with
two (constrained) free parameters, namely $\alpha_3 \leq 0$ and $\chi \in
\mathbb{R} \backslash 0$. The proportionality factor $\chi$ is in general
different for all $\lambda^{(n)}$ and can in principle be calculated from higher
order vertex functions. In our truncation we choose a uniform constant for
simplicity. The \IR{} structure is unaffected by the value of $\chi$, but a
large $\chi$ might alter the \UV{} regime. Note that the scaling analysis is
true in the \IR{} limit only. Consequently, we expect a small $\chi$ which
does not interfere with the \UV{} regime. 

\section{Anomalous dimensions in the \IR}
\label{appendix_etaIR}

Both anomalous dimensions need to vanish at the \IR{} fixed point for
it to be classical. For the ghost anomalous dimension, this is the case as long
as $g/\epsilon \rightarrow 0$, which is equivalent to saying that $\alpha_3 < 
0$,
in accordance with our estimate above. On the other hand,
the vanishing of the graviton anomalous dimension is seen as follows:
First, as shown in \cite{Christiansen:2012rx}, the terms quadratic in the
 external momentum in the flow cancel. Thus, the flow goes to a constant as the
 external momentum goes to infinity, and no divergences can appear there. Next,
 the flow equation for the mass can be rewritten as
 \begin{equation}
  \dot\mu = -2\mu + \frac{\partial_t \Gamma^{(2h)}(0)}{Z_h(0)} + \eta_h(0) \mu
 \, .
 \end{equation}
 We know that in the \IR{} for $\mu \to -1, g\to 0$, $\dot\mu$ vanishes. This
 can be achieved in 3 different ways: either the flow vanishes and the anomalous
 dimension at zero cancels the canonical scaling, or both the flow and the
 anomalous dimension cancel the canonical scaling, or only the flow remains
 finite. First assume that the flow vanishes in the limit $\mu \to -1, g \to 0$.
 We know that the leading order contribution comes from a term $\sim g
 \lambda^{(4)}$, all other terms have smaller divergences and thus vanish in the
 limit $g \to 0$. If this term vanishes, however, then the flow is 0 everywhere,
 thus also the anomalous dimension would vanish everywhere, and we would end up
 with no fixed point. On the other hand, assume that $g \lambda^{(4)}$ remains
 finite in our limit, then still all other terms in the flow vanish, and thus
 the flow is a nonzero constant. This in turn implies that the graviton
 anomalous dimension must vanish as it is a finite difference of the flow. We
 conclude that $\eta_h(p^2)=0$ at the \IR{} fixed point.

\bibliography{flatgravity}

\end{document}